\documentclass[pdflatex,sn-mathphys-num]{sn-jnl}

\usepackage{tabularray}
\UseTblrLibrary{booktabs}
\usepackage{multirow}
\usepackage{graphicx}%
\usepackage{multirow}%
\usepackage{amsmath,amssymb,amsfonts}%
\usepackage{amsthm}%
\usepackage{mathrsfs}%
\usepackage[title]{appendix}%
\usepackage{xcolor}%
\usepackage{textcomp}%
\usepackage{manyfoot}%
\usepackage{booktabs}%
\usepackage{algorithm}%
\usepackage{algorithmicx}%
\usepackage{algpseudocode}%
\usepackage{listings}%
\usepackage{overpic}
\usepackage{braket}
\usepackage{mathrsfs}
\usepackage{float}
\usepackage{multirow}
\usepackage{array}
\usepackage{booktabs}
\usepackage[utf8]{inputenc}
\usepackage{hhline}
\usepackage{makecell}

\theoremstyle{thmstyleone}%
\theoremstyle{thmstyletwo}%

\theoremstyle{thmstylethree}%

\begin{document}

\title[Article Title]{Fundamentals and Applications of Hybrid Electro- and Opto- mechanical system coupled to Superconducting Qubit: A Short Review}


\author[1]{\fnm{Roson} \sur{Nongthombam}}\email{n.roson@iitg.ac.in}
\equalcont{These authors contributed equally to this work.}

\author[1]{\fnm{Urmimala } \sur{Dewan}}\email{d.urmimala@iitg.ac.in}
\equalcont{These authors contributed equally to this work.}

\author*[1]{\fnm{Amarendra K. } \sur{Sarma}}\email{aksarma@iitg.ac.in}
\equalcont{These authors contributed equally to this work.}

\affil[1]{\orgdiv{Department of Physics}, \orgname{Indian Institute of Technology Guwahati}, \orgaddress{ \city{Guwahati}, \postcode{781039}, \state{Assam}, \country{India}}}


\abstract{Superconducting qubits, realized by incorporating Josephson junctions into superconducting circuits, behave as artificial atoms with anharmonic energy spectra and can be precisely controlled and measured using microwave cavities within the framework of circuit quantum electrodynamics (cQED). Since its emergence in the early 2000s, cQED has established superconducting qubits as leading candidates for scalable quantum devices and has enabled the exploration of hybrid quantum systems that integrate disparate physical platforms. This review surveys superconducting hybrid quantum electromechanical systems in which mechanical resonators are coupled to superconducting qubits, with a focus on two widely used qubit platforms: the transmon and the fluxonium. We provide an overview of the underlying coupling mechanisms arising from interactions through the phase and charge degrees of freedom of the qubit, and discuss how these mechanisms give rise to both longitudinal and transverse qubit–mechanical interactions. We further review extensions of electromechanical platforms to electro-optomechanical architectures, in which optical cavities are integrated to enable coherent interfacing between superconducting circuits and optical photons. This review aims to present a unified framework and perspective on qubit–mechanical and qubit–mechanical–optical hybrid systems in superconducting quantum technologies and sensor applications.}





\maketitle
\section{\label{sec:Introduction}Introduction}
The past few decades have witnessed remarkable advances in quantum information science, including quantum computation~\cite{ref1}, quantum communication~\cite{Bennett1992}, quantum cryptography~\cite{Lo2014}, and quantum sensing~\cite{RevModPhys.89.035002,10.1063/5.0339361}, paving the way toward a new quantum era. Quantum sensing, in particular, capitalizes on the intrinsic sensitivity of quantum systems to external disturbances to achieve high-precision measurements of physical quantities, ranging from weak-field detection to rotation and pressure, with emerging applications in gravitational wave detection \cite{Harry_2010} and dark matter detection \cite{Carney_2021}. The foundations of quantum computation were laid when Richard Feynman and others introduced the idea of a 'universal computer' at the 1981 summit, followed by famous papers in 1982 and 1985 \cite{feynman1,Feynman1985}, such as ``Simulating Physics with Quantum Computers." As Feynman put it, "The problem is, how can we simulate quantum mechanics? ... Can you do it with a new kind of computer — a quantum computer? Now it turns out, as far as I can tell, that you can simulate this with a quantum system, with quantum computer elements. It's not a Turing machine, but a machine of a different kind." As quantum computation theory progressed \cite{Deutsch,simon,shor1,shor2}, the need for physically realizable qubits emerged. Several platforms were explored to implement qubits such as nuclear spins (NMR)\cite{NMR}, electronic spins of trapped ions and quantum dots\cite{PhysRevLett.74.4091,PhysRevA.57.120}, and photon polarization\cite{Knill2001}. The idea of superconducting qubits originated based on the foundational work on Josephson junction and superconducting quantum interference device \cite{josephson1962,Jaklevic1964,Jaklevic1965}, unlocking the potential for quantum coherence in superconducting circuits \cite{cal1983,martinis1985}. 
Unlike natural atoms, superconducting circuits offer distinct advantages because their transition frequencies and other parameters can be precisely engineered. 
Over the past two decades, this flexibility has led to the development of several superconducting qubit architectures, including flux qubits \cite{fluxqubit1}, phase qubits \cite{phasequbit1,phasequbit2}, and improved charge-based designs such as the transmon qubit \cite{transmon1}. 
Each qubit type comes with its own set of challenges; for example, charge qubits are particularly sensitive to charge noise, whereas flux and phase qubits are predominantly affected by flux noise. 
The evolution from early qubit implementations to more robust designs with enhanced coherence times has been extensively reviewed in the literature \cite{Clarke2008,Devoret2004}.

The development of superconducting qubits, motivated by cavity quantum electrodynamics and its exploration of photon–atom interactions, led to the emergence of the circuit QED paradigm.
Circuit QED \cite{RevModPhys.93.025005,cqed1,PhysRevA.75.032329}, which is analogous to cavity QED, studies microwave photons coupled to superconducting qubits acting as artificial atoms. 
The first demonstration of cQED came from Wallraff et al. \cite{wallraff} in 2004, when they experimentally showed strong coupling of superconducting qubits with a microwave photon in a transmission line resonator (a superconducting cavity), effectively realizing circuit QED and paving the way for the development of a modified charge qubit - the transmon qubit. 
The idea of transmon was first introduced by Koch. et al. \cite{transmon1} as a charge-noise-insensitive qubit. 
Another example in the same direction is the fluxonium qubit \cite{manucharyan2009fluxonium,PRXLee,PhysRevX.11.011010,nguyen2022blueprint,ding2023high,PhysRevLett.129.010502,PhysRevX.9.041041}, where both flux and charge noises can be suppressed by choosing appropriate parameters. 

Additionally, cQED has facilitated the development of hybrid quantum systems that combine distinct physical platforms within a single architecture \cite{clerk2020}. These hybrid systems leverage the complementary strengths of the constituent components, enabling enhanced functionality, tunable interactions, and new regimes of quantum control and measurement that are not accessible with individual systems alone.
The aim of hybrid quantum systems is to achieve coherent coupling between cQED modes—such as microwave cavity photons and the quantum state of a superconducting qubit—and fundamentally different physical excitations \cite{PhysRevLett.105.140501}, including mechanical phonons \cite{OConnell2010}, microwave photons \cite{wallraff}, and spin-wave modes in ferromagnets (magnons) \cite{tabuchi,Petersson2012}. 

A notable and rapidly developing hybrid cQED platform is the quantum electromechanical system. 
In these systems, various mechanical resonators
can be utilized, each offering distinct coupling mechanisms with qubits.
Mechanical resonators provide a versatile platform for quantum hybrid architectures. 
Structures such as suspended membranes, cantilevers, nanobeams, and drumheads, beyond their conventional use in classical technologies, open pathways to the quantum domain, where they couple coherently to superconducting qubits through charge or phase degrees of freedom. 
In these systems, suspended micro- and nanomechanical resonators operating at megahertz frequencies are coupled to superconducting qubits primarily in two ways. The coupling can be realized either capacitively, by replacing one of the qubit's capacitor plates with a suspended mechanical element \cite{Pirkkalainen2013,Lecocq2015,LaHaye2009}, or inductively, by allowing one arm of the superconducting loop to move out of the plane \cite{Bera2021,Bera2024,Rodrigues2019}. Qubits can also couple to surface or bulk acoustic modes confined in phononic crystals, which form low-loss, high-quality phononic resonators \cite{doi:10.1126/science.aaw8415,doi:10.1126/science.aao1511,OConnell2010}. Such coupling is also commonly implemented using a flip-chip architecture, with the qubit and resonator fabricated on separate chips \cite{PRXLee,Wollack2022,Chu2018}.
Qubit–mechanical oscillator systems combine the high force sensitivity of mechanical resonators with the precise control and high-fidelity readout of qubits \cite{10.1063/5.0021088}. In circuit quantum electrodynamics (cQED), such systems are widely used for electromechanical sensing, where mechanical motion is transduced into measurable signals via superconducting circuits \cite{PhysRevA.111.013509, suri2025}.

The mechanical resonator in an electromechanical system can be made to interact with an optical resonator, forming a cavity optomechanical system. At the core of cavity optomechanics \cite{RevModPhys.86.1391} is radiation pressure, a concept first postulated by Kepler in the 17th century, later formalized by Maxwell, and experimentally demonstrated in 1901 \cite{lebedew,PhysRevSeriesI.13.307}. The earliest experimental prototypes of cavity optomechanical systems date back to the late 1970s, when Russian physicist Braginsky and co-workers, motivated primarily by the development of early-stage gravitational wave detectors, constructed large-scale interferometers \cite{braginski1967,braginsky1968}. Following two decades of progress \cite{PhysRevLett.45.75,braginsky1995quantum}, the quantum aspects of optomechanical systems began to be explored in the 1990s \cite{PhysRevA.47.3173,Brooks2012}. These studies include squeezing of light \cite{PhysRevA.50.4237,PhysRevA.49.1337}, quantum nondemolition detection of light intensity \cite{PhysRevA.49.1961,Heidmann1997}, and feedback cooling of mechanical oscillators \cite{PhysRevLett.80.688}. Furthermore, by exploiting strong optomechanical coupling, researchers highlighted the possibility of generating nonclassical and entangled states of light and mechanical motion \cite{PhysRevA.56.4175}.
The basic principle of sensing in optomechanical systems relies on the coupling between light and mechanical motion, where small displacements of a mechanical oscillator modify the optical field. 
This interaction enables precise measurements of forces, displacements, and fields. 
As a result, optomechanical systems have become a powerful platform for sensing \cite{Rossi2018}, with applications including the measurement of acceleration, mass, acoustic signals, displacement \cite{PhysRevLett.122.071101}, and weak forces \cite{PhysRevLett.121.031101,doi:10.1126/science.1249850}. However, their performance is ultimately limited by noise; shot noise and back-action noise play a crucial
role in determining the sensitivity limit \cite{RevModPhys.52.341}.

A coupled qubit–mechanical–optical system forms a hybrid electromechanical system, which can be realized by introducing an optical channel into an electromechanical setup. For example, a suspended nano- or micromechanical resonator can couple to an optical mode by forming a Fabry-Pérot optomechanical cavity, where the resonator serves as one of the cavity mirrors \cite{PhysRevA.94.012340,andrews2014bidirectional,groeblacher2015observation}. Alternatively, optomechanical crystals \cite{Chan2011,Barzanjeh2022,Cohen2015,PhysRevLett.121.220404} can couple a qubit to the breathing mechanical mode of the crystal via an interdigital transducer (IDT), which generates surface acoustic waves (SAWs) on either piezoelectric or non-piezoelectric substrates using a piezoelectric film \cite{Balram2016,Mirhosseini2020,Jiang2020,10.1063/1.4955408}.
Such hybrid systems are particularly valuable for quantum transduction, a key requirement for long-distance quantum communication and the construction of large-scale quantum networks. They are also instrumental for ground-state cooling of mechanical resonators \cite{Arcizet2006,PhysRevA.104.023509}.

This review article is structured as follows. Section \ref{sec: Fundamentals of Superconducting Circuits} introduces the basic concepts of a two-level system, a simple harmonic oscillator, and provides a brief introduction to superconductivity. These concepts form the foundation for constructing hybrid quantum systems, which are explored in the subsequent sections.
In Section \ref{sec:Transmon and Fluxonium}, we present a brief introduction to the Cooper pair box (CPB) and discuss how its different operating regimes give rise to various qubits, namely, the charge qubit and the transmon qubit. The section also describes the implementation of the fluxonium qubit and explains how these qubits can be read out using a microwave cavity in the dispersive regime.
Section \ref{sec:Qubit-Mechanical hybrid system} focuses on the qubit–mechanical hybrid system, exploring how a mechanical resonator couples to the phase and charge degrees of freedom of the qubit. This section also discusses relevant implementations and potential applications.
In Section \ref{sec:Optomechanical Coupling with Qubit}, we extend this discussion to the qubit–mechanical–optical hybrid system, highlighting its implementation and applications. Finally, Section \ref{Conclusion} presents the concluding remarks.

\section{ Basic Concepts}
\label{sec: Fundamentals of Superconducting Circuits}

The prominence of superconducting qubits has increased significantly due to their central role in quantum computing architectures. Notably, several leading quantum processors developed by companies such as \(\text{Google}\) and \(\text{IBM}\) are based on superconducting circuits \cite{Arute2019,doi:10.1126/science.aao4309,Wang2018}. From a physical perspective, superconducting qubits function as effective two-level systems, making their study closely connected to the broader framework of quantum optics and two-level quantum systems. While an in-depth understanding of superconductivity is not strictly required, a basic familiarity with its underlying principles is advantageous.
In this section, we briefly review the properties of two-level systems, followed by a discussion of the quantum harmonic oscillator and a concise introduction to superconductivity. We then present the essential concepts of superconducting circuits required for the remainder of this review.

\subsection{Two-State Quantum Systems}
A two-level system is the simplest model in quantum optics~\cite{gerry2023introductory}. It consists of a ground state $\lvert g\rangle$ and an excited state $\lvert e\rangle$. An arbitrary state of such a system can be written as
\begin{equation}
\lvert \psi\rangle = a_g \lvert g \rangle + a_e \lvert e\rangle ,
\end{equation}
where $a_g$ and $a_e$ are the probability amplitudes associated with the ground and excited states, respectively. The energies of the excited and ground states are taken to be $\epsilon_z$ and $-\epsilon_z$.

The most general Hamiltonian of a two-level quantum system can be expressed in terms of the Pauli matrices as,
 $\hat{\sigma}_{x} = \begin{pmatrix}
    0 & 1 \\
    1 & 0
\end{pmatrix}$ , $\hat{\sigma}_{y} = \begin{pmatrix}
    0 & -i \\
    i & 0
\end{pmatrix}$ and $\hat{\sigma}_{z} = \begin{pmatrix}
    1 & 0 \\
    0 & -1
\end{pmatrix}$, as
\begin{equation}
\hat{H} = \epsilon_x \hat{\sigma}_x + \epsilon_y \hat{\sigma}_y + \epsilon_z \hat{\sigma}_z .
\end{equation}
If the two-level system (or atom) is not coupled to any external field, the off-diagonal terms vanish, and the Hamiltonian reduces to a diagonal form describing a bare two-level system.

\subsection{Harmonic Oscillator}

We briefly recall the quantum mechanics of the harmonic oscillator. In classical mechanics, the harmonic oscillator is commonly modeled as a mass–spring system, in which a mass $m$ is attached to a spring with spring constant $k$. The dynamics of this system provide the starting point for its quantum mechanical description.
The harmonic oscillator satisfies the equation of motion \(m\ddot{x}=-kx=-m\omega^2x\), where \(\omega\) is the oscillator frequency. Its classical Hamiltonian is \(H=\frac{p^2}{2m}+\frac{1}{2}m\omega^2x^2\), with \(x\) and \(p\) forming a pair of canonically conjugate variables. Upon quantization, the Hamiltonian becomes \(\hat H=\frac{\hat p^2}{2m}+\frac{1}{2}m\omega^2\hat x^2\). Introducing the ladder operators through \(\hat x=x_{\rm ZPF}(\hat a+\hat a^\dagger)\) and \(\hat p=i m\omega x_{\rm ZPF}(\hat a^\dagger-\hat a)\), where \(x_{\rm ZPF}=\sqrt{\hbar/(2m\omega)}\) is the zero-point fluctuation amplitude, the Hamiltonian can be written as \(\hat H=\hbar\omega(\hat a^\dagger\hat a+\tfrac{1}{2})\). The annihilation and creation operators satisfy \(\hat a|n\rangle=\sqrt{n}\,|n-1\rangle\) and \(\hat a^\dagger|n\rangle=\sqrt{n+1}\,|n+1\rangle\), where \(|n\rangle\) denotes the Fock state. These states are eigenstates of the harmonic oscillator Hamiltonian, \(\hat H|n\rangle=E_n|n\rangle\), with eigenenergies \(E_n=\hbar\omega(n+\tfrac{1}{2})\) for \(n=0,1,2,\ldots\).

Due to unavoidable coupling to a thermal environment, observing quantum behavior in an oscillator is challenging, as thermal noise broadens discrete energy levels into a classical continuum. Quantum effects become accessible only when the energy-level spacing exceeds both the linewidth (dephasing) and the thermal energy $k_B T$. The former requires a high-quality factor $Q=\omega/\kappa$, with values up to $10^8$ achievable in superconducting circuits ~\cite{wallraff}, while the latter is ensured by cooling to $k_B T \ll \hbar\omega$, readily attainable in dilution refrigerators operating below $20\,\mathrm{mK}$~\cite{wallraff}.
A defining feature of a quantum harmonic oscillator is its equally spaced energy levels, separated by $\hbar\omega$. Realizing an effective two-level system, therefore, requires breaking this harmonic ladder by introducing anharmonicity. In a superconducting LC circuit (quantum harmonic oscillator), this is achieved by replacing the linear inductor with a Josephson junction, which consists of two superconducting islands separated by a thin insulating barrier and acts as a nonlinear inductive element. This Josephson junction~\cite{josephson1962} forms the fundamental building block of superconducting circuits.

\subsection{Basics of Superconductivity}

For completeness, we briefly review the essential aspects of superconductivity, the physical phenomenon that underlies superconducting circuits and Josephson-junction-based quantum devices.


For ordinary metals, the electrical resistance decreases with decreasing temperature but never vanishes completely. In contrast, a superconductor exhibits an abrupt transition to zero resistance at a critical temperature, $T_c$. This phenomenon of superconductivity was discovered by Heike Kamerlingh Onnes in 1911~\cite{Onnes1911a,Onnes1911b,Onnes1911c}, for which he was awarded the Nobel Prize in Physics in 1913. Another defining property of superconductors is their response to magnetic fields. When cooled below $T_c$, superconductors expel applied magnetic fields from their interior, thereby exhibiting perfect diamagnetism. This effect, discovered in 1933 by Walther Meissner and Robert Ochsenfeld, is known as the Meissner effect~\cite{MeissnerOchsenfeld1933}.
 
Microscopic superconductivity is described by the BCS (Bardeen-Cooper-Schrieffer) theory~\cite{PhysRev.108.1175,tinkham2004introduction}. Within this framework, electrons form bound pairs known as Cooper pairs through phonon-mediated interactions. Each Cooper pair carries charge $2e$ and condenses into a macroscopic quantum state characterized by long-range phase coherence. A key result of BCS theory is the explicit form of the superconducting ground-state wave function, which describes a many-electron state exhibiting long-range phase coherence. The Cooper pairs in this state share a common macroscopic phase, well defined across the entire system, giving rise to a macroscopic quantum state known as the BCS condensate. At temperatures below the critical temperature $T_c$, the superconductor remains close to this coherent ground state. The energy required to break a Cooper pair defines the superconducting energy gap $2\Delta$, which at zero temperature satisfies the BCS relation $2\Delta(0)\approx 3.5\,k_B T_c$. The emergence of the BCS condensate enables dissipationless charge transport, as elastic scattering processes cannot degrade the supercurrent.

An alternative and complementary phenomenological description is provided by the Ginzburg-Landau theory~\cite{ginzburg2009theory,RevModPhys.76.981}, which introduces a complex order parameter $\psi(\vec{r})$ to characterize the superconducting condensate. Within this framework, the order parameter can be written as
$
\psi(\vec{r}) = \sqrt{\rho(\vec{r})}\,e^{i\theta(\vec{r})},
$
where $\rho(\vec{r}) = |\psi(\vec{r})|^2$ is proportional to the local Cooper-pair density and $\theta(\vec{r})$ denotes the macroscopic superconducting phase.
A direct consequence of the single-valuedness of the phase of the order parameter is the quantization of the magnetic flux $\Phi$ threading a superconducting loop:
$
\Phi = n\frac{h}{Q},
\label{eq:flux_quantization}
$
where $n$ is an integer and $Q=2e$. The flux can therefore be written as $\Phi=n\Phi_0$, where $\Phi_0=\frac{h}{2e}$ is the superconducting flux quantum.

A superconducting loop containing one or more Josephson junctions is formed by interrupting a superconducting ring with thin insulating barriers. In such a loop, the flux quantization condition constrains the superconducting phase differences across the junctions. The current flowing through a Josephson junction is given by $I = I_c \sin(\Delta\theta)$, where $\Delta\theta$ is the superconducting phase difference across the junction and $I_c$ is the critical current, defined as the maximum supercurrent that can flow through the junction before losing its superconducting state. The phase difference $\Delta\theta$ determines the Josephson potential energy and therefore serves as the relevant dynamical degree of freedom. Any externally applied magnetic flux—including flux generated by mechanical motion-modulates the Josephson potential, thereby coupling the superconducting phase to other degrees of freedom, such as the motion of a mechanical resonator. This mechanism provides a fundamental basis for realizing phase-dependent interactions in hybrid superconducting systems.
\section{Superconducting qubits: Transmon and Fluxonium}
\label{sec:Transmon and Fluxonium}
\subsection{Cooper Pair Box (CPB)}
\begin{figure}[htbp]
    \centering
    \includegraphics[width=85mm]{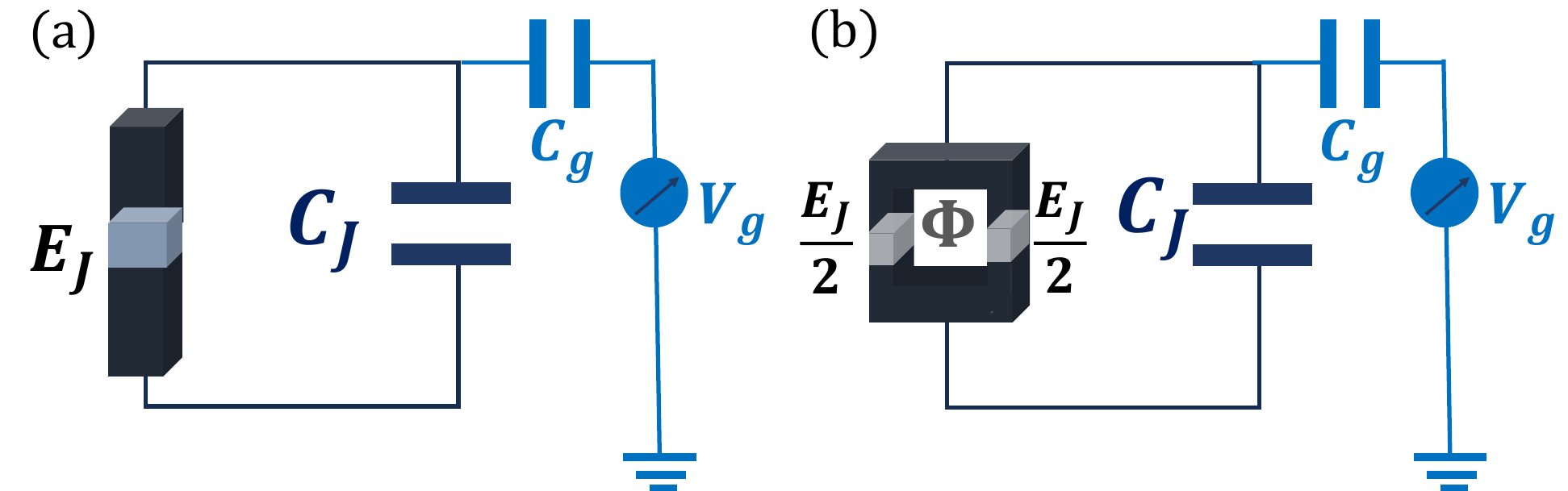}
    \caption{
    (a) Circuit diagram of a Cooper pair box (CPB). It consists of two superconducting electrodes separated by a junction and electrostatically biased by a voltage applied through a small capacitor.  
    (b) Circuit diagram of a split Cooper pair box (CPB), where the two superconducting electrodes are connected by two junctions.
    }
    \label{fig:system_transmon}
\end{figure}
In this section, we provide an overview of superconducting qubit implementations based on Josephson junction circuits.
We discuss the Cooper pair box and its evolution into distinct operating regimes—namely, the charge, transmon, and fluxonium regimes, each reflecting a different balance between charging and Josephson energies.

The fact that the quantum state of a superconductor below a certain critical temperature is described by a single, non-degenerate ground-state wavefunction can be leveraged to engineer an artificial atom at the mesoscopic level. 
For this purpose, we require another non-degenerate ground state to which Cooper pairs can tunnel. 
This can be achieved by placing two superconductors next to each other, separated by a small insulating junction, which is referred to as a Josephson junction. 
Cooper pairs can then tunnel from one superconductor to the other through the junction, which is essential for realizing the artificial atom.
The Hamiltonian of this two superconducting island setup can be written by noting the following:
(a) When a Cooper pair tunnels from one superconductor to the other, the upper island becomes positively charged and the lower island becomes negatively charged, causing the system to behave like a capacitor with capacitance $C_J$ and giving rise to electrostatic energy.
(b) There is a tunnelling energy term in the Hamiltonian that allows changes in the particle number between the two superconducting islands.

Now, consider two metallic superconducting electrodes connected by an insulating Josephson junction (JJ). The electrodes are in their ground states, where the electrons in each electrode are fully paired. 
The total number of electron pairs in the system is fixed.
Through the junction, electron pairs can tunnel from the upper to the lower superconductor. The tunnel junction allows for the coherent tunnelling of electron pairs from one side to the other. 
This coherent transfer is described by the Hamiltonian,
\begin{equation}
\label{HJ}
    \hat{H}_J = -\frac{1}{2}E_J \sum_m \left( |m\rangle\langle m+1| + |m+1\rangle\langle m| \right),
\end{equation}
where \( |m\rangle \) represents the quantum state corresponding to \( m \) Cooper pairs having tunneled across the junction, and \( E_J \) is the Josephson coupling energy.
The two superconductors, along with the junction in between, can be electrostatically biased by a voltage coupled through a small capacitor, as shown in Fig. \ref{fig:system_transmon}(a). This setup forms a Cooper pair box (CPB).
Let $\hat{n} = \sum_m m\, |m\rangle\langle m|$
be the charge operator that counts the number of Cooper pairs transferred across the junction. The charging energy of the CPB is then given by
\begin{equation}
\label{HC}
    \hat{H}_C = 4E_C (\hat{n} - n_g)^2,
\end{equation}
where \( E_C = \frac{e^2}{2(C_J + C_g)} \) is the charging energy, and \( n_g = C_g V_g / 2e \) is the dimensionless gate charge controlled by the gate voltage \( V_g \).
The full Hamiltonian describing the dynamics of the CPB is
\begin{equation}
\label{HJC}
    \hat{H} = 4E_C (\hat{n} - n_g)^2 - \frac{1}{2}E_J \sum_m \left( |m\rangle\langle m+1| + |m+1\rangle\langle m| \right).
\end{equation}
The charge operator can be expressed in terms of the superconducting phase difference \( \theta \) across the junction using the relations:
\begin{eqnarray}
\label{n_theta}
    \hat{n} &=& -i\frac{\partial}{\partial\theta}, \\ 
    |\theta\rangle &=& \sum_{n=-\infty}^{\infty} e^{i n \theta} |n\rangle,  \\ 
    |n\rangle &=& \frac{1}{2\pi} \int_0^{2\pi} d\theta\, e^{-i n \theta} |\theta\rangle.
\end{eqnarray}
\begin{figure}[htbp]
    \centering
    \includegraphics[width=85mm]{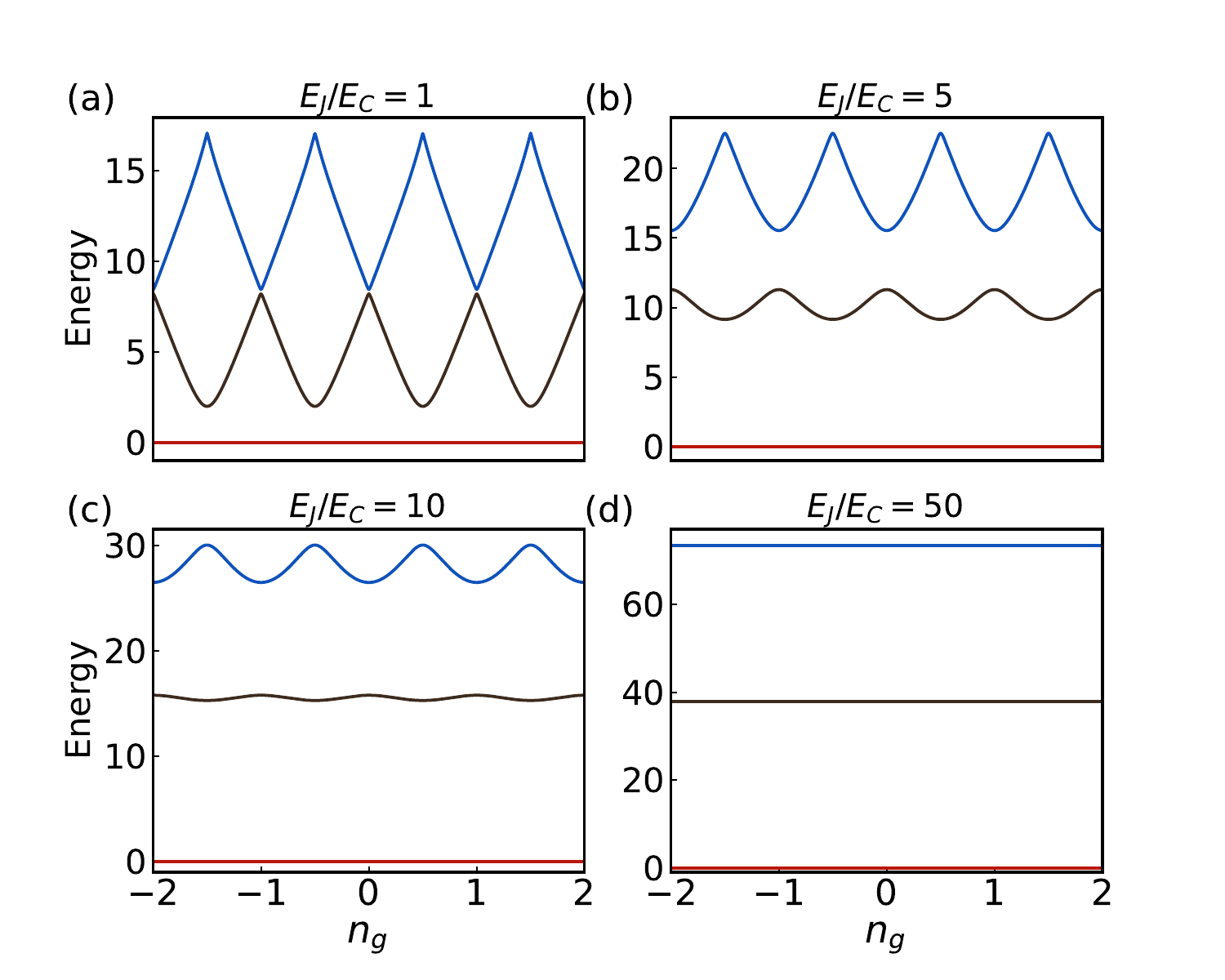}
    \caption{Energy spectrum of the qubit Hamiltonian as a function of the gate charge $n_{g}$ 
    for different values of the ratio \(E_J^{\text{eff}}(\Phi_e)/4E_C \). With an increasing ratio, the charge dispersion is progressively suppressed, and the energy levels become effectively insensitive to $n_{g}$.}
    \label{fig:transmon energy}
\end{figure}
The phase states \( |\theta\rangle \) form a complete, orthonormal basis over the interval \( [0, 2\pi] \), and are related to the charge (number) states \( |n\rangle \) via Fourier transform.
Using these relations, the Hamiltonian in Eq.~\eqref{HJC} can be written in the phase basis as
\begin{equation}
\label{H_phase}
    \hat{H} = 4E_C \left(-i\frac{\partial}{\partial \theta} - n_g\right)^2 - E_J {\cos(\hat{\theta})},
\end{equation}
where \( \theta \) is the conjugate variable to the number operator \( \hat{n} \), and the cosine term arises from the Josephson tunneling of Cooper pairs.

Instead of a single junction, one can realize two Josephson junctions between the two superconducting electrodes, forming a split Cooper pair box (CPB), as shown in Fig. \ref{fig:system_transmon}(b). 
The loop formed by this split junction is also referred to as a SQUID loop.
The Josephson energy of the split CPB is given by \cite{RevModPhys.93.025005}
\begin{equation}
\label{H_split}
    \hat{H}_{SJ} = -E_{J1} \cos(\hat{\theta}_1) - E_{J2} \cos(\hat{\theta}_2),
\end{equation}
where \( E_{J1} \) and \( E_{J2} \) are the Josephson energies of the two junctions, and \( \hat{\theta}_1 \), \( \hat{\theta}_2 \) are the superconducting phase differences across each junction.
If an external magnetic flux \( \Phi_e \) is applied through the loop, then by the condition of flux quantization, we have \(\hat{\theta}_1 - \hat{\theta}_2 = 2\pi \Phi_e/\Phi_0,\) where \( \Phi_0 = h/2e \) is the superconducting flux quantum \cite{transmon1}.
Introducing a new collective phase variable \(\hat{\theta} = (\hat{\theta}_1 + \hat{\theta}_2)/2,\)
the Hamiltonian in Eq.~\eqref{H_split} becomes
\begin{equation}
    \hat{H}_{SJ} = -E_J^{\text{eff}}(\Phi_e) \cos(\hat{\theta}),
\end{equation}
where the effective Josephson energy is flux-dependent:
\begin{equation}
    E_J^{\text{eff}}(\Phi_e) = E_J^{\text{max}} \left[ \cos^2\left( \frac{\pi \Phi_e}{\Phi_0} \right) + d^2 \sin^2\left( \frac{\pi \Phi_e}{\Phi_0} \right) \right]^{1/2},
\end{equation}
with\[
E_J^{\text{max}} = E_{J1} + E_{J2}, \quad d = \frac{E_{J1} - E_{J2}}{E_{J1} + E_{J2}}.\]

Returning to the charge basis, the Hamiltonian of the split Cooper pair box (CPB) can be written as
\begin{equation}
\label{CPB}
\begin{aligned}
\hat{H} ={}& \sum_m 4E_C (m - n_g)^2 \ket{m}\bra{m} \\
&- \frac{1}{2} E_J^{\text{eff}}(\Phi_e)
\sum_m \left( \ket{m}\bra{m+1} + \ket{m+1}\bra{m} \right),
\end{aligned}
\end{equation}
where \( E_J^{\text{eff}}(\Phi_e) \) is the flux-tunable Josephson energy obtained from the split-junction configuration.
This Hamiltonian can be diagonalized numerically to obtain the energy eigenvalues \( E_i(\Phi_e) = \langle i| \hat{H} |i\rangle \) and the corresponding eigenstates \( |i\rangle \). The diagonalized form of the Hamiltonian is
\begin{equation}
\label{CPB_diag}
    \hat{H}_{\text{CPB}} = \sum_i E_i(\Phi_e) |i\rangle\langle i|.
\end{equation}
By varying the external magnetic flux \( \Phi_e \), the energy spectrum of the CPB can be tuned. 
The energy levels \( E_i(\Phi_e) \) are, in general, anharmonic, meaning that the spacing between adjacent levels is not uniform. 
This anharmonicity is crucial for defining a two-level system (qubit), since it allows selective excitation of the lowest transition \( |0\rangle \leftrightarrow |1\rangle \) without exciting higher levels.
The degree of anharmonicity depends on the ratio \( E_J^{\text{eff}}(\Phi_e)/4E_C \) as shown in Fig. \ref{fig:transmon energy}. 
When this ratio is small, the anharmonicity is large and one can form a charge qubit from this CPB regime. When this ratio is large, the CPB enters the so-called transmon regime.

\subsection{Charge Regime}
\begin{figure}[htbp]
    \centering
    \includegraphics[width=80mm]{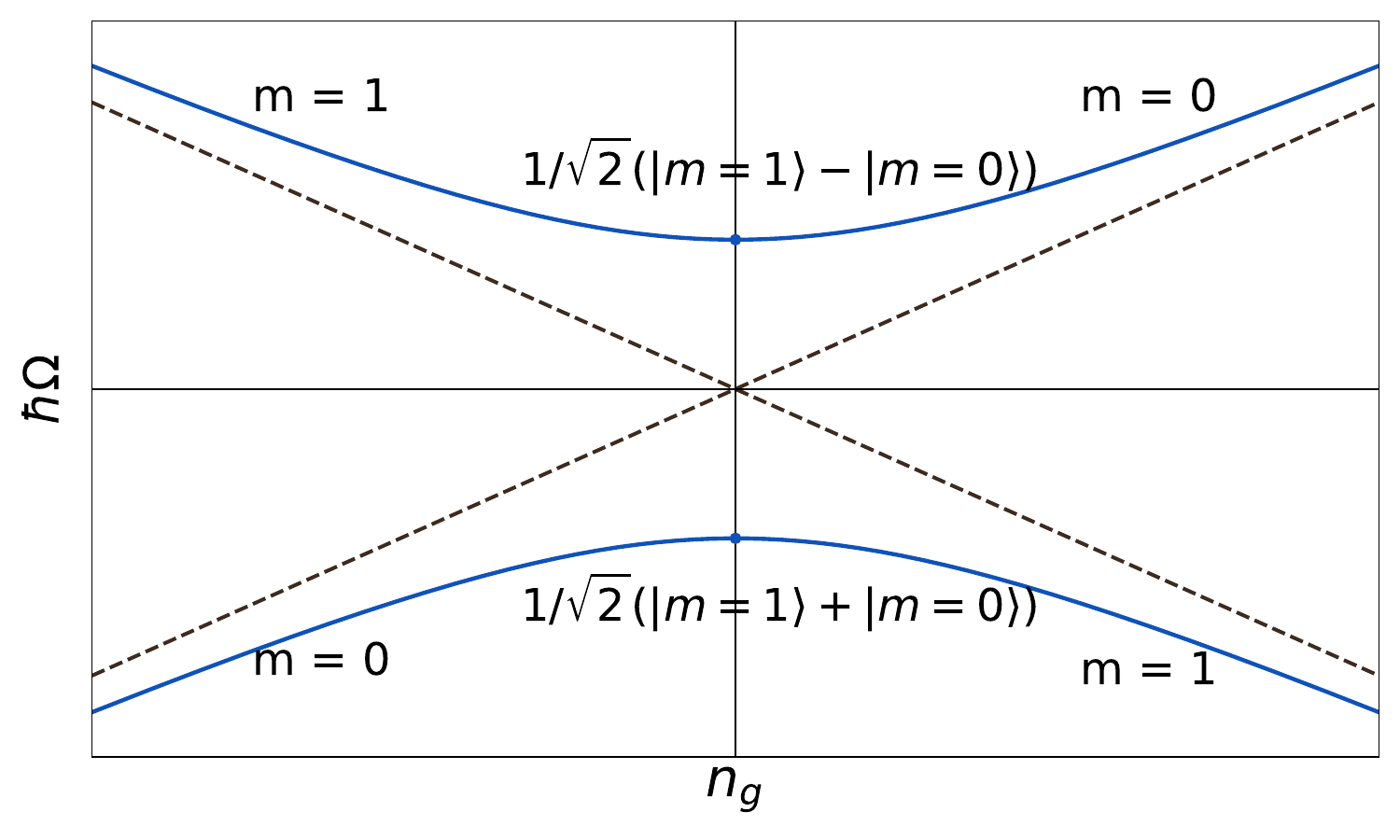}
    \caption{Energy spectrum of the charge qubit as a function of the gate charge $n_{g}$. The dotted lines correspond to the bare energies of the charge states in the absence of Josephson coupling, $(E_{J} = 0)$. The solid lines represent the dressed eigenstates obtained after diagonalizing the qubit Hamiltonian. At $n_{g} = \frac{1}{2}$, an avoided crossing is clearly observed due to the Josephson tunnelling of Cooper pairs.}
    \label{fig:eigenvalue_charge}
\end{figure}

In this regime of the CPB, the Josephson tunnelling energy is dominated by the charging energy from the biased voltage, i.e., \( E_J^{\text{eff}}(\Phi_e)/4E_C < 1\). 
As a result, as the gate charge $n_g$ changes, the energy spectrum also changes significantly. 
This leads to charge induced fluctuations in the transition energy. 
At $n_g = 1/2$, however, the transition energy is insensitive, to first order, to small fluctuations of $n_g$.
The Hamiltonian for a two-level system or qubit in this charge regime can be obtained from Eq. \eqref{CPB} by considering $n_g\in [0,1]$ and taking only up to the first tunnelling Cooper pairs $(m=0,1)$. By neglecting the resultant offset terms, one obtains
\begin{equation}
    \label{eqn:charge_qubit}
    \hat{H}_c = - \frac{E_{el}}{2} \, \hat{\sigma}_z - \frac{E_J^{\text{eff}}(\Phi_e)}{2} \, \hat{\sigma}_x,
\end{equation}
where $E_{el}= 4E_C(1-2n_g) $. By transforming into the qubit basis, the Hamiltonian Eq. \eqref{eqn:charge_qubit} can be written as
\begin{equation}
    \label{eqn:cq1}
    \hat{H}_q = \frac{\hbar\Omega}{2}\sigma_z,
\end{equation}
where $\hbar\Omega=(E_J^2 + (4E_C(1-2n_g))^2)^{1/2}$. Here, the transformation is $\sigma_z \rightarrow -\sigma_z\cos\theta + \sigma_x\sin\theta$ and $\sigma_x \rightarrow-\sigma_x\cos\theta  -\sigma_z \sin\theta$, where $\tan\theta = E_J^{\text{eff}}(\Phi_e)/E_{el}$.
The eigenvalue spectrum of the charge qubit is shown in Fig. \ref{fig:eigenvalue_charge}.

\subsection{Transmon Regime}

\begin{figure}[htbp]
    \centering
    \begin{overpic}[width=63mm]{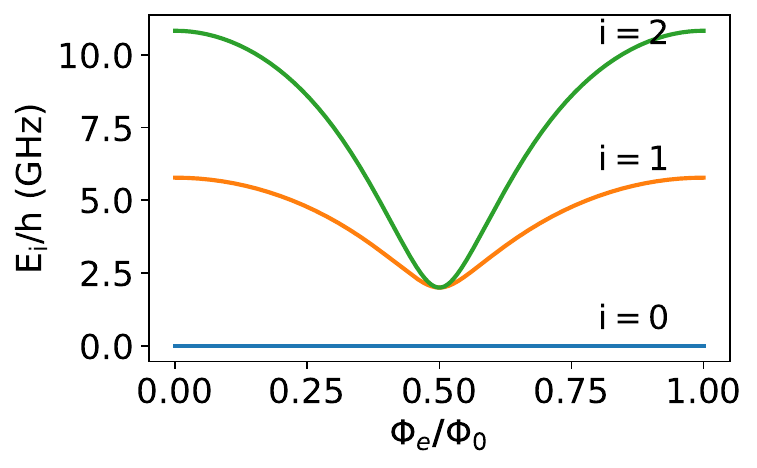}
        \put(-4,55){ (a)}
    \end{overpic}
    \hspace{3mm}
    \begin{overpic}[width=61mm]{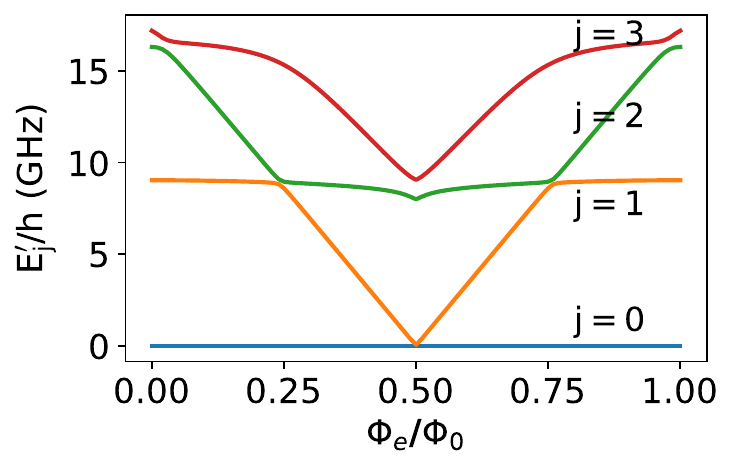}
        \put(-4,55){ (b)}
    \end{overpic}

    \vspace{-2mm}
    \caption{Eigenvalue spectrum of (a) Transmon and (b) Fluxonium with respect to the externally applied flux $\Phi_e/\Phi_0$. The indices $i$ and $j$ denote the energy eigenstates, with $i = 0,1,2,\ldots$ corresponding to the transmon and $j = 0,1,2,3$ corresponding to the fluxonium.
    Parameters used: (a) $E_{J1}=E_{J2} = 5 \, \text{GHz}$ and $E_C=0.5 \, \text{GHz}$, (b) $E_J = 10 \, \text{GHz}$,  $E_C = 1.2\, \text{GHz}$, and $E_L =  1\, \text{GHz}$.}
    \label{fig:trans_spec}
\end{figure}

When the ratio \( E_J^{\text{eff}}(\Phi_e)/4E_C \) is large, the CPB is in the transmon regime, where the anharmonicity of the energy levels decreases, as shown in Fig. \ref{fig:transmon energy}. 
In the transmon regime ($E_J/E_C \gg 1$), the charge dispersion is exponentially suppressed, rendering the qubit insensitive to charge noise. Although this comes at the expense of reduced anharmonicity, the charge dispersion decreases exponentially with $(8E_J/E_C)^{-1/2}$, whereas the anharmonicity decreases only algebraically as $(8E_J/E_C)^{-1/2}$ \cite{transmon1}. Consequently, the dependence of the Hamiltonian in Eqs.~\eqref{CPB} and \eqref{CPB_diag} on the offset charge $n_g$ can be neglected.
The energy spectrum can still be changed by modifying the externally applied field $\Phi_e$.
The corresponding energy spectrum is shown in Fig. \ref{fig:trans_spec}(a), which exhibits weak but non-negligible anharmonicity that is sufficient for qubit operation while offering enhanced coherence properties. 
It should be noted that the energy spectrum can also be derived from the phase-basis Hamiltonian using Mathieu functions, where the Schrödinger equation takes the form of a Mathieu differential equation \cite{transmon1,10.1063/1.446581,milne1972handbook}. This approach provides analytical insight into the energy levels.

Instead of evaluating the transmon energy spectrum exactly by numerical diagonalization of the Hamiltonian in Eq.~\eqref{CPB_diag}, one can employ a perturbative approach to obtain an analytical approximation.
Since the anharmonicity is small in the transmon regime, we retain only the leading nonlinear contribution from the potential term $-E_J^{\mathrm{eff}}(\Phi_e)\cos\hat{\theta}$ in the Hamiltonian of Eq. ~\eqref{CPB}, so that the transmon Hamiltonian reads
\begin{equation}
    \label{eqn:transmon}
    \hat{H}_T = 4E_C\hat{n}^2 + E_J^{eff}(\Phi_e)\left(-1 + \frac{1}{2}\hat{\theta}^2-\frac{1}{4!}\hat{\theta}^4 \right)
\end{equation}
The terms containing $\hat{n}^2$ and $\hat{\theta}^2$ together constitute a harmonic oscillator. 
The Hamiltonian of this oscillator can be written in terms of the annihilation and creation operators 
$\hat{b}$ and $\hat{b}^\dagger$, defined as
\begin{align}
    \label{eqn:tranihilatio}
    \hat{\theta} =& \left(\frac{2E_C}{E_J^{eff}(\Phi_e)}\right)^{1/4}(\hat{b}+\hat{b}^{\dagger}),\nonumber\\
    \hat{n} =& \frac{i}{2}\left(\frac{E_J^{eff}(\Phi_e)}{2E_C}\right)^{1/4}(\hat{b}^{\dagger}-\hat{b}).
\end{align}
Treating the harmonic oscillator as the unperturbed Hamiltonian and the remaining terms as a perturbation, 
we obtain from Eq.~\eqref{eqn:transmon}
\begin{align}
    \label{eqn:transmon1}
    \hat{H}_T \approx  \, \hbar\Omega_T (\hat{b}^\dagger\hat{b}+\frac{1}{2})-E_J^{eff}(\Phi_e)-\frac{E_C}{12}(6(\hat{b}^\dagger\hat{b})^2+6 \, \hat{b}^\dagger\hat{b}+3),
\end{align}
where $\hbar\Omega_T = \sqrt{8E_C E_J^{eff}(\Phi_e)}$ is the Josephson plasma frequency. In the second term of Eq. \eqref{eqn:transmon1}, the rotating wave approximation (RWA) is applied, which is valid here since $\hbar\Omega_T~>>~E_C/4$. 
The operators $\hat{b}^\dagger$ and $\hat{b}$ constitute a simple harmonic approximation to the transmon qubit. 
In terms of this basis, the eigenenergies of the transmon become
\begin{equation}
    \label{eqn:eigentrans}
    E_n =  \hbar\Omega_T (n+\frac{1}{2})-E_J^{eff}(\Phi_e)-\frac{E_C}{12}(6n^2+6 \, n+3)
\end{equation}
Using this relation, we can calculate the anharmonicity of the transmon $\alpha=(E_2-E_1)-(E_1-E_0)$ and relative anharmonicity $\alpha_r=\alpha/(E_1-E_0)$. We obtain $\alpha=-E_C$ and $\alpha_r=-(8E_J^{eff}(\Phi_e)/E_C)^{-1/2}$. 
In circuit Quantum Electrodynamics (QED) experiments, transmon qubits are designed with transition frequencies typically in the gigahertz (GHz) range. 
The anharmonicity of a transmon, given by $E_C/h$,  is in the range of a few hundred megahertz (MHz). 
Although the anharmonicity is much smaller than the transition frequency, it is sufficiently large compared to the linewidth of the energy levels, making it possible to resolve the anharmonicity experimentally. 
This ability to resolve anharmonicity is crucial in circuit QED experiments, as it allows for precise control over the qubit energy levels and interactions.

\subsection{Fluxonium}
\begin{figure}[htbp]
    \centering
    \includegraphics[width=65mm]{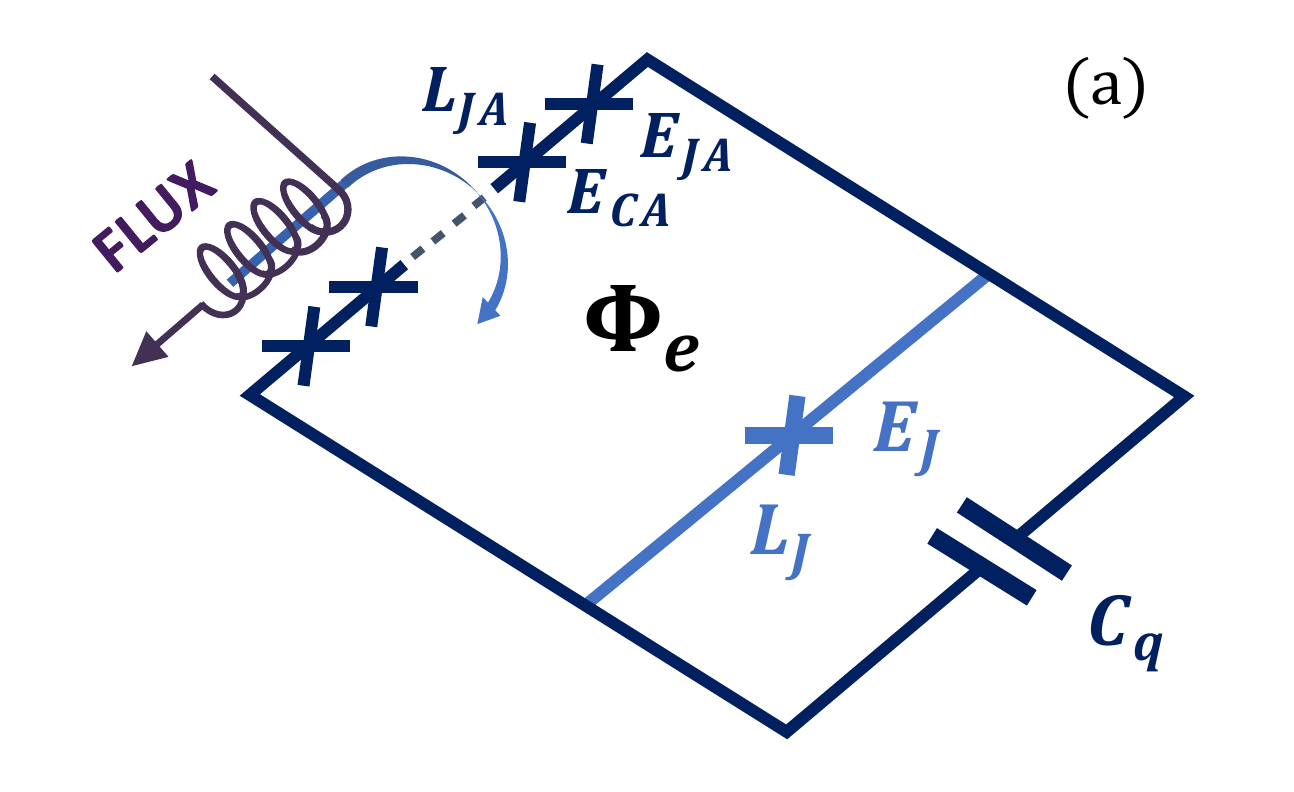}
    \vspace{-30mm} 
    \includegraphics[width=75mm]{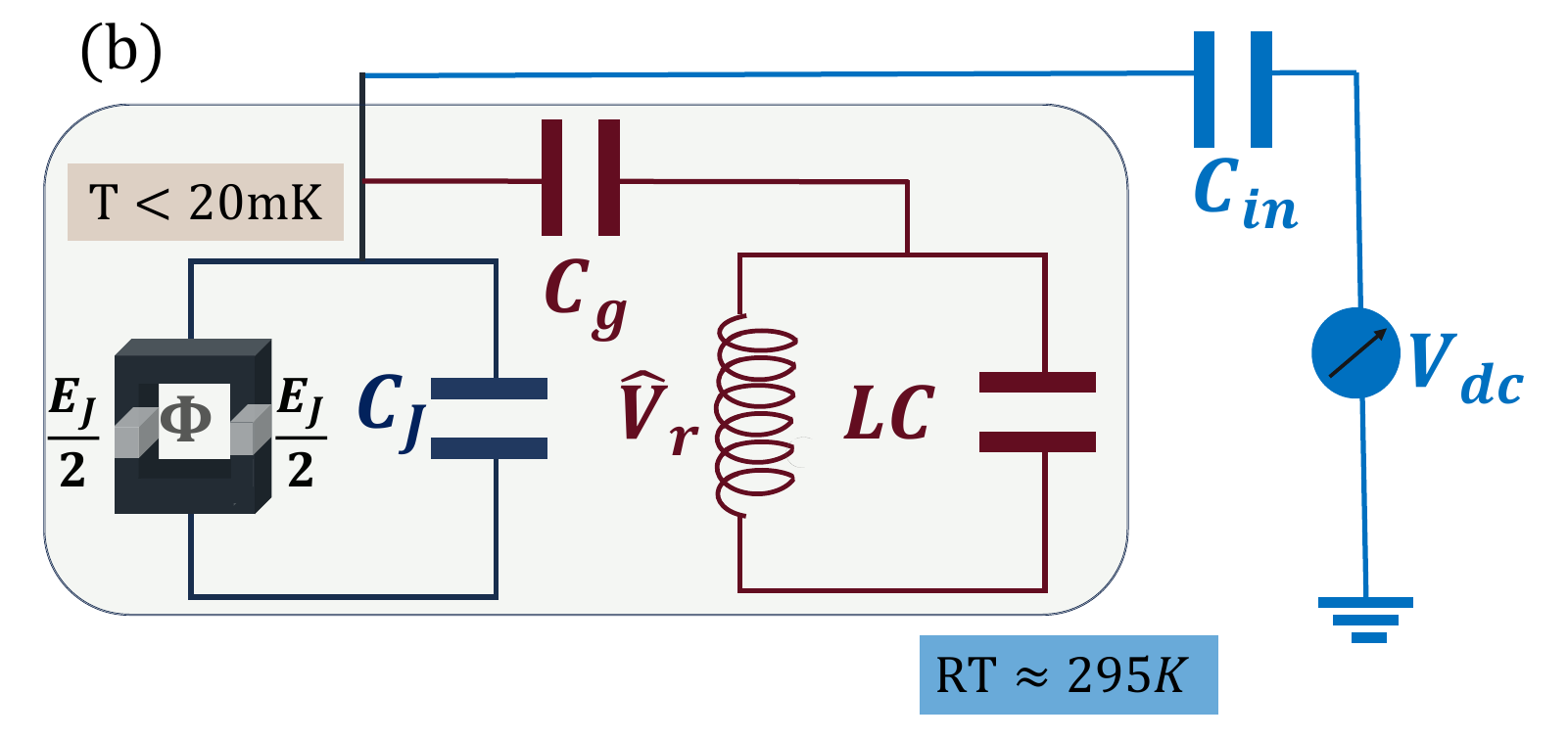}
    \vspace{30mm}
    \caption{
    (a) Circuit diagram of a fluxonium. The fluxonium qubit consists of a Josephson junction array shunted by a single junction and a capacitor. The flux loop can be biased by an external flux applied perpendicular to the loop. The symbol $\times$ represents a Josephson junction. Adapted from \cite{PhysRevX.11.011010}.  (b) Schematic of qubit–cavity coupling. A microwave cavity, represented by an LC oscillator, is coupled to the qubit through the capacitor $C_g$. Additionally, a DC bias is connected to the qubit via an input capacitor $C_{\text{in}}$. The qubit–cavity system is maintained in a cold environment (below $20~\text{mK}$) to preserve quantum behaviour, while the external DC bias is applied at room temperature ($T_R$).
    }
    \label{fig:system_fluxonium}
\end{figure}

 Fluxonium is realized similarly to the split Cooper pair box (CPB), but with one key difference: one of the Josephson junctions is replaced by an array of approximately $N=300$ large-area Josephson junctions, each with Josephson energy \( E_{JA} \) and charging energy \( E_{CA} \). We assume \( E_{JA}/E_{CA} \gg 1 \), ensuring that the charge dispersion in each array junction is small, as shown in Fig. \ref{fig:system_fluxonium}(a)\cite{manucharyan2009fluxonium,PhysRevX.14.011007,PhysRevX.11.011010}.
The resulting array acts as a super-inductance that effectively shunts the remaining small Josephson junction, contributing an inductance \( L_{JA} \) and a capacitance \( C_q \). The Josephson energy of the entire array is given by
\[
\sum_{i=1}^{N} -E_{JA} \cos\left( \frac{\hat{\theta}}{N} \right) \approx \frac{E_{JA}}{2N} \hat{\theta}^2,
\]
assuming \( \hat{\theta}/N \ll 1 \). 
Here, \( \theta \) is the total superconducting phase difference across the \textit{small} Josephson junction, and \( N \) is the number of junctions in the array. The phase drop across each junction in the array is \( \hat{\theta}/N \) due to flux quantization. The junctions in the chain are much larger than the small junction characterized by Josephson energy \( E_J \), and they serve to suppress quantum phase slips that can lead to decoherence.
If an external magnetic flux \( \Phi_e \) is applied through the loop, the Hamiltonian for the fluxonium circuit becomes
\begin{equation}
    \hat{H}_F = 4 E'_C \hat{n}^2 - E_J \cos\left( \hat{\theta} + 2\pi \frac{\Phi_e}{\Phi_0} \right) + \frac{1}{2} E_L \hat{\theta}^2,
\end{equation}
where \( E'_C = \frac{e^2}{2 C_q} \) is the charging energy, and \( E_L = \frac{E_{JA}}{N} \) is the effective inductive energy from the array. The operators \( \hat{\theta} \) and \( \hat{n} \) are conjugate variables representing the phase and number of Cooper pairs across the small junction.
Taking the first and third terms as a quantum harmonic oscillator (QHO) Hamiltonian, the system can be diagonalized in the harmonic oscillator basis. This yields
\begin{equation}
    \hat{H}_{\text{F}} = \sum_j E'_j(\Phi_e) |j\rangle\langle j|,
\end{equation}
where \( E_j(\Phi_e) = \langle j| \hat{H}_F |j\rangle \) are the energy eigenvalues and \( |j\rangle \) are the corresponding eigenstates.
By varying the external magnetic flux \( \Phi_e \), the energy spectrum of the fluxonium circuit can be tuned, as shown in Fig. \ref{fig:trans_spec}(b). 
As in the CPB case, the energy levels \( E_j(\Phi_e) \) are anharmonic, allowing the implementation of a qubit. 
The anharmonicity in fluxonium is significantly richer than that in the transmon regime because the fluxonium eigenstates contain substantial contributions from many harmonic oscillator states, unlike the transmon where only adjacent levels dominate. 
This richer structure provides increased flexibility for qubit design and control.

\subsection{Dispersive Readout}

The states of both transmon and fluxonium qubits can be read out by coupling them to a readout cavity.
The qubit is placed inside a microwave resonator and capacitively coupled to the resonator mode. The microwave resonator can be either a two-dimensional planar waveguide or a three-dimensional cavity in which the electromagnetic field is confined \cite{10.1063/1.3010859,Simons2001CPW,Pozar2005Microwave,Day2003KID, PhysRevLett.107.240501}.
In both types of resonators, the qubit is positioned at the centre, where the antinode of the confined electromagnetic field is located. 
Although the resonators support multiple modes, we focus only on the mode with a frequency that is close to the qubit frequency, neglecting all other modes by assuming that they are far detuned from the qubit.
The configuration for the transmon readout is shown in Fig.~\ref{fig:system_fluxonium}(b), and a similar configuration applies to the fluxonium.
Since the coupling is capacitive, the cavity mode interacts with the qubit through the charging term in the qubit Hamiltonian.
More specifically, the charge operator $\hat{n}$ is replaced by $\hat{n} + \hat{n}_r$, where the cavity mode acts as an effective offset charge for the qubit. Substituting this into the transmon and fluxonium Hamiltonians yields the composite cavity–qubit Hamiltonians:
\begin{align}
\hat{H}^{\mathrm{T}}_{\mathrm{C}} 
&= 4 E_C^{\mathrm{T}} \bigl(\hat{n}^{\mathrm{T}}\bigr)^2
 - E_J^{\mathrm{eff}}(\Phi_e)\cos\!\bigl(\hat{\theta}^{\mathrm{T}}\bigr)
 + \hbar g^{\mathrm{T}} (\hat{a}^\dagger + \hat{a}) \hat{n}^{\mathrm{T}} 
 + \hbar \omega \hat{a}^\dagger \hat{a}, \\[6pt]
\hat{H}^{\mathrm{F}}_{\mathrm{C}}
&= 4 E_C^{\mathrm{F}} \bigl(\hat{n}^{\mathrm{F}}\bigr)^2
 - E_J \cos\!\left( \hat{\theta}^{\mathrm{F}} + 2\pi \frac{\Phi_e}{\Phi_0} \right)
 + \frac{1}{2} E_L \bigl(\hat{\theta}^{\mathrm{F}}\bigr)^2 
 + \hbar g^{\mathrm{F}} (\hat{a}^\dagger + \hat{a}) \hat{n}^{\mathrm{F}}
 + \hbar \omega \hat{a}^\dagger \hat{a},
\end{align}
where the coupling constants are $g^{\text{T}} = 2E_C^{\text{T}} (C_g^{\text{T}}/e)x_{\text{zpf}}$ and $g^{\text{F}} = 2E_C^{\text{F}}(C_g^{\text{F}}/e) x_{\text{zpf}}$, and $x_{\text{zpf}}$ is the vacuum fluctuation of the cavity mode.
Here, $\hat{\theta}^{\text{T(F)}}$ and $\hat{n}^{\text{T(F)}}$ are the superconducting phase and the corresponding number operator for the transmon (fluxonium), respectively.
We have assumed that each qubit interacts with only one mode of the cavity whose frequency is close to the qubit transition frequency.
The above Hamiltonians can be rewritten in the uncoupled transmon and fluxonium basis states $|i_{\text{T}}\rangle$ and $|j_{\text{F}}\rangle$:
\begin{align}
    \hat{H}^{\text{T}}_{\text{C}} &= \sum_i E_i^{\text{T}} |i_{\text{T}}\rangle\langle i_{\text{T}}| + \hbar g^{\text{T}}\,(\hat{a}^\dagger + \hat{a})\,\sum_{i,l} \langle i_{\text{T}}|\hat{n}^{\text{T}}|l_{\text{T}}\rangle\,|i_{\text{T}}\rangle\langle l_{\text{T}}| 
    + \hbar\omega\,\hat{a}^\dagger\hat{a}, \\
    \hat{H}^{\text{F}}_{\text{C}} &= \sum_j E_j^{\text{F}} |j_{\text{F}}\rangle\langle j_{\text{F}}| + \hbar g^{\text{F}}\,(\hat{a}^\dagger + \hat{a})\,\sum_{j,k} \langle j_{\text{F}}|\hat{n}^{\text{F}}|k_{\text{F}}\rangle\,|j_{\text{F}}\rangle\langle k_{\text{F}}| 
    + \hbar\omega\,\hat{a}^\dagger\hat{a},
\end{align}
The dispersive interaction can be obtained by performing the Schrieffer–Wolff transformation \cite{PhysRevB.87.024510,RevModPhys.93.025005}. The dispersive Hamiltonians then read:
\begin{align}
\label{dispersive}
    \hat{H}^{\text{T}}_{\text{C}} &= \sum_i (E_i^{\text{T}} + \eta_i^{\text{T}}) |i_{\text{T}}\rangle\langle i_{\text{T}}| + \hbar \,\hat{a}^\dagger\hat{a}\,\sum_i \chi_i^{\text{T}} |i_{\text{T}}\rangle\langle i_{\text{T}}| + \hbar\omega\,\hat{a}^\dagger\hat{a}, \\
    \hat{H}^{\text{F}}_{\text{C}} &= \sum_j (E_j^{\text{F}} + \eta_j^{\text{F}}) |j_{\text{F}}\rangle\langle j_{\text{F}}| + \hbar \,\hat{a}^\dagger\hat{a}\,\sum_j \chi_j^{\text{F}} |j_{\text{F}}\rangle\langle j_{\text{F}}| + \hbar\omega\,\hat{a}^\dagger\hat{a}.
\end{align}
Here, the Lamb shifts and dispersive shifts are given by:
\begin{align}
    \eta_i^{\text{T}} &= \sum_l \chi_{il}^{\text{T}}, \quad \chi_i^{\text{T}} = \sum_l (\chi_{il}^{\text{T}} - \chi_{li}^{\text{T}}), \\
    \eta_j^{\text{F}} &= \sum_k \chi_{jk}^{\text{F}}, \quad \chi_j^{\text{F}} = \sum_k (\chi_{jk}^{\text{F}} - \chi_{kj}^{\text{F}})
\end{align}
with
\begin{align}
    \chi_{il}^{\text{T}} &= \frac{\hbar (g^{\text{T}})^2 |\langle i_{\text{T}}|\hat{n}^{\text{T}}|l_{\text{T}}\rangle|^2}{E_i^{\text{T}} - E_l^{\text{T}} - \hbar\omega}, \quad
    \chi_{jk}^{\text{F}} = \frac{\hbar (g^{\text{F}})^2|\langle j_{\text{F}}|\hat{n}^{\text{F}}|k_{\text{F}}\rangle|^2}{E_j^{\text{F}} - E_k^{\text{F}} - \hbar\omega}.
\end{align}
From the Hamiltonians in Eq.~\eqref{dispersive}, it is clear that the frequency of the cavity is shifted by an amount $\chi_i^{\text{T}}$ or $\chi_j^{\text{F}}$, depending on the state $i$ or $j$ of the transmon or fluxonium qubit, respectively. 
Measuring this shift allows one to infer the state of the qubit.
In the case of the transmon, the eigenstates can be approximated by those of a harmonic oscillator, and the transition matrix elements $\chi_{il}^{\text{T}}$ are non-zero primarily for adjacent levels, and asymptotically approach zero for higher transitions. 
Under the rotating wave approximation, the dispersive shift can be expressed as,
\begin{equation}
    \chi_{i,i+1}^{\text{T}} = \frac{\hbar (g^{\text{T}})^2 |n_0|^2(i+1)}{E_i^{\text{T}} - E_{i+1}^{\text{T}} - \hbar\omega},
\end{equation}
where $n_0 = \frac{i}{\sqrt2} \left( \frac{E_J^{\text{eff}}}{8E_C^{\text{T}}} \right)^{\frac{1}{4}}$ represents the vacuum fluctuation of the charge operator. From this relation, both the Lamb shift and the dispersive shift of the qubit can be calculated.
In contrast, for fluxonium qubits, the eigenstates cannot be approximated by those of a harmonic oscillator, as higher Fock state contributions are significant. 
However, the dispersive shift can still be computed numerically.

\section{Qubit-Mechanical hybrid system}
\label{sec:Qubit-Mechanical hybrid system}

Both the fluxonium and transmon qubits possess charge and phase degrees of freedom—channels through which they can be controlled and coupled to other systems.
An external degree of freedom, such as a mechanical resonator can interact with these qubits via either channel. 
We first focus on coupling through the charge channel and address the flux channel later.

\subsection{Coupling through the charge degree of freedom}
\begin{figure*}[htbp]
    \centering
    \includegraphics[width=125mm]{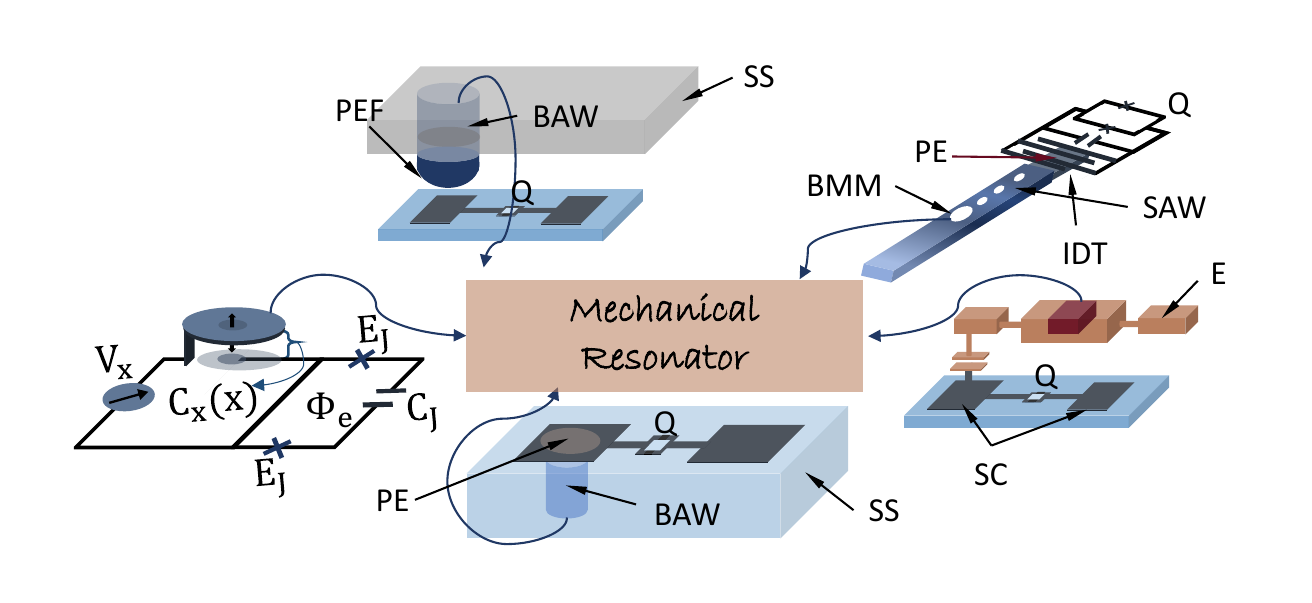}
    \caption{
    Schematic of a hybrid electromechanical system.  
    A suspended mechanical resonator is coupled to the qubit by changing the capacitance $C_x(x)$ due to mechanical motion (bottom left).  
    In the centre, two bulk acoustic wave (BAW) phononic crystals are coupled to the qubit (Q), either using a flip-chip architecture (top) or on a single-chip sapphire substrate (denoted by SS) (bottom). In both cases, the electric field from the qubit excites the piezoelectric film (PEF), typically AlN, which in turn excites the BAW modes in the substrate.  
    On the top right, a patterned mechanical crystal is coupled to the qubit through a piezoelectric slab (PES). The excitation from the qubit is first transferred to the PES via an interdigitated transducer (IDT), and then to a breathing mechanical mode (BMM) in the crystal through surface acoustic waves (SAW).  
    Another realization of the flip-chip architecture is shown on the bottom right, where the superconducting (SC) qubit is coupled to a piezoelectric phononic mode (PPM) deposited on the defects of the electrode (E). Adapted from
    \cite{doi:10.1126/science.aao1511, Mirhosseini2020, 10.1063/5.0021088, clerk2020}}
    \label{fig:system_qubit_mech}
\end{figure*}
\begin{figure}[htbp]
    \centering
    \includegraphics[width=85mm]{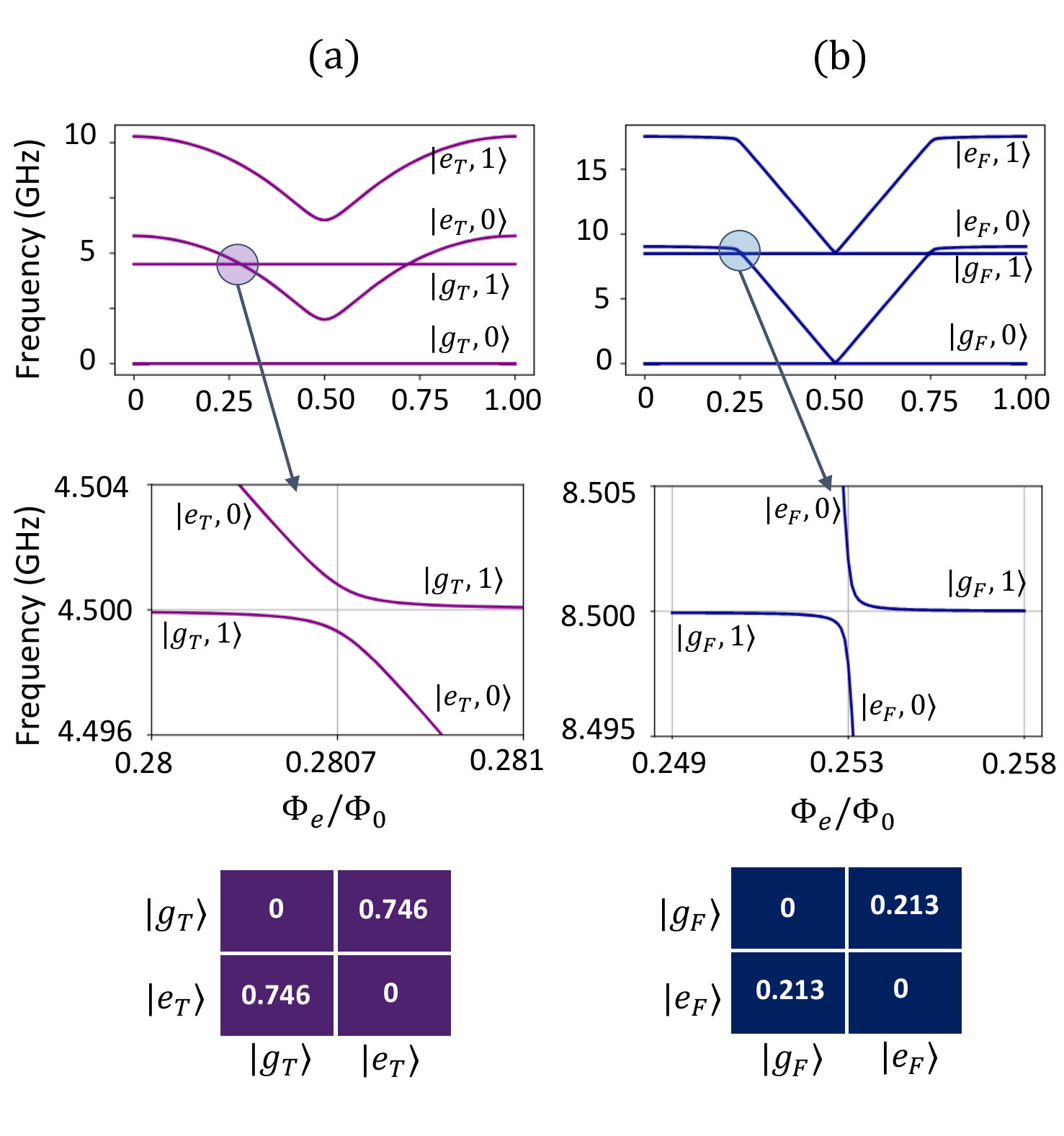}
    \caption{Avoided crossing energy spectrum of the (a) transmon qubit–mechanical and (b) fluxonium qubit–mechanical coupled systems. The charge matrix elements $n_{ij}^{\text{T}}$ and $n_{ij}^{\text{F}}$, where $i,j = e,g$, at the external flux point where the avoided crossing occurs are shown below. Parameters used: (a) $E_{J\text{max}} = 10\,\text{GHz}$, $E_C = 0.5\,\text{GHz}$, $d = 0$, $g_T = 1\,\text{MHz}$, $\omega_b = 4.5\,\text{GHz}$; (b) $E_J = 10\,\text{GHz}$, $E_C = 1.2\,\text{GHz}$, $E_L = 1\,\text{GHz}$, $g_F = 10\,\text{MHz}$, $\omega_b = 8.5\,\text{GHz}$.}

    \label{fig:transverse_avoided_spec}
\end{figure}

Coupling mediated through the charge degree of freedom occurs via capacitive interaction between the resonator and the qubit. 
Depending on the type of mechanical resonator, this coupling can arise either from the piezoelectric response of the resonator material or from the motion-dependent capacitance of the capacitor that connects the resonator to the qubit \cite{Gerashchenko2025,Pirkkalainen2013} as shown in Fig. \ref{fig:system_qubit_mech}.
In the case of piezoelectric coupling, the resonator is typically made from a material such as aluminum nitride (AlN) with a phononic crystal structure \cite{OConnell2010}. 
The electric field generated by the qubit excites acoustic modes in the resonator through the material's intrinsic piezoelectricity. 
The excited mode can be either a bulk acoustic wave (BAW) or a surface acoustic wave (SAW). For SAW-based devices, this excitation is facilitated by an interdigitated transducer (IDT), which converts the qubit’s electric signal into propagating surface waves \cite{Gustafsson2014, Noguchi2017}.
In both cases, the excited acoustic mode can be effectively modelled as an offset to the qubit's charge operator in the Hamiltonian, i.e., 
$\hat{n} \rightarrow \hat{n} + \hat{n}_r$
where $\hat{n}_r$ is proportional to the resonator’s zero-point displacement and typically takes the form $\hat{n}_r \propto \hat{b} + \hat{b}^\dagger$ with $\hat{b}$ the annihilation operator of the mechanical mode.
In the transmon case, under capacitive coupling to the mechanical resonator, the charge operator is modified to $\hat{n}_T \rightarrow \hat{n}_T + \hat{n}_r$
Projecting onto the qubit subspace spanned by $\{\langle g_T|, \langle e_T|\}$, the effective Hamiltonian  becomes:

\begin{align}
\hat{H}_{\text{TM}} &= \hbar \omega_b \hat{b}^\dagger \hat{b} +\sum_{i = g_T, e_T} E_i^{\text{T}} |i_T\rangle\langle i_T| 
+\hbar g_{\text{T}} \sum_{i,j = g_T, e_T} n_{ij}^{\text{T}} |i_T\rangle\langle j_T| \left( \hat{b} + \hat{b}^\dagger \right)
\label{eq:H_trans_mech}
\end{align}
where, $n_{ij}^{\text{T}} = \langle i_T| \hat{n}_T |j_T\rangle$ are the transmon charge matrix elements, $g_{\text{T}} = \dfrac{4 E_C^{\text{T}} C^{\text{T}}_m}{\hbar ~e} x_0$ is the transmon–mechanical coupling strength, $x_0$ is the zero-point motion amplitude of the resonator.
Similarly, in the fluxonium case, coupling the resonator capacitively yields $\hat{n}_F \rightarrow \hat{n}_F + \hat{n}_r$
Projecting onto the low-energy fluxonium qubit subspace $\{\langle g_F|, \langle e_F|\}$ gives:

\begin{align}
\hat{H}_{\text{FM}} = \hbar \omega_b \hat{b}^\dagger \hat{b} + \sum_{i = g_F, e_F} E_i^{\text{F}} |i_F\rangle\langle i_F| + \hbar g_{\text{F}} \sum_{i,j = g_F, e_F} n_{ij}^{\text{F}} |i_F\rangle\langle j_F| \left( \hat{b} + \hat{b}^\dagger \right),
\label{eq:H_FM}
\end{align}

where, $n_{ij}^{\text{F}} = \langle i_F| \hat{n}_F |j_F\rangle$ are the fluxonium charge matrix elements, $g_{\text{F}} = \dfrac{4 E_C^{\text{F}} C^\text{F}_m}{\hbar ~ e} x_0$ is the fluxonium–mechanical coupling strength. 
By diagonalizing Eq. \eqref{eq:trans_mech} and Eq. \eqref{eq:H_FM}, the avoided crossing spectrum of the transmon-mechanical and fluxonium-mechanical coupled systems can be obtained which is shown in Fig. \ref{fig:transverse_avoided_spec}.
Some experimental realizations of transmon and fluxonium qubits coupled to mechanical resonators through the charge degree of freedom are shown in Fig.~\ref{fig:table}. These include suspended mechanical drums, nanomechanical beams, and mechanical acoustic resonators such as BAW and SAW devices, as well as flip-chip architectures.

In terms of Pauli matrices, the Hamiltonian $H_{\text{TM}}$ and $H_{\text{FM}}$ can be written as,

\begin{eqnarray}
    \hat{H}_{\text{TM}} &=& \frac{1}{2} \hbar \omega_\text{0T} \hat{\sigma}_z +
    \hbar g_{\text{T}} \, 
    n_{ge}^{\text{T}}\left[ \hat{\sigma}_+ + \hat{\sigma}_-
    \right]
    \left( \hat{b} + \hat{b}^\dagger \right), \\
    \hat{H}_{\text{FM}} &=& \frac{1}{2} \hbar \omega_\text{0F} \hat{\sigma}_z +
    \hbar g_{\text{F}} \, n_{eg}^{\text{F}}\left[
    \hat{\sigma}_+ + \hat{\sigma}_-
    \right]
    \left( \hat{b} + \hat{b}^\dagger \right).
\end{eqnarray}
where, $\omega_\text{0T} = \frac{1}{\hbar}(E_e^{\text{T}} - E_g^{\text{T}})$ and $\omega_\text{0F} = \frac{1}{\hbar}(E_e^{\text{F}} - E_g^{\text{F}})$. 
The longitudinal components are ignored, since their matrix elements are zero. 
Both Hamiltonians have the same structure and resemble the Jaynes–Cummings Hamiltonian.
To analyse the Jaynes–Cummings interaction, we omit the T and F labels for simplicity, as the analysis applies equally to both transmon and fluxonium qubits.
The full Jaynes–Cummings Hamiltonian under RWA is:

\begin{equation}
\hat{H} = \hat{H}_0 + \hat{H}_I = \hbar \omega_b \hat{b}^\dagger \hat{b} + \frac{1}{2} \hbar \omega_q \hat{\sigma}_z + \hbar G \left( \hat{\sigma}_+ \hat{b} + \hat{\sigma}_- \hat{b}^\dagger \right).
\end{equation}

We define the detuning as $\Delta = \omega_q - \omega_b$. The coupling constant G is dependent on the matrix element $n^\text{T,F}_{eg}$ Because the Jaynes-Cummings Hamiltonian conserves the total number of excitations, defined by the excitation number operator: $\hat{N} = b^\dagger b + \sigma_+ \sigma_-$ the Hamiltonian acts within invariant subspaces of fixed excitation number \( N = n + 1 \). 
Each subspace has the basis: $\left\{ |e, n\rangle,\; |g, n+1\rangle\right\}$
Thus, the total Hamiltonian in the \( \{ |e,n\rangle, |g,n+1\rangle \} \) basis becomes:
\begin{equation}
    H_n = \hbar \omega_b (n+\frac{1}{2}) \, I +
    \hbar
    \begin{pmatrix}
    \frac{\Delta}{2} & G \sqrt{n+1} \\
    G \sqrt{n+1} & -\frac{\Delta}{2}
    \end{pmatrix}
\end{equation}
The eigenvalues are:
\begin{equation}
    E_n^{(\pm)} = \hbar \omega_b (n+\frac{1}{2}) \pm \hbar \Omega_n, \quad \text{where } \Omega_n = \sqrt{G^2(n+1) + \left( \frac{\Delta}{2} \right)^2}.
\end{equation}
and the corresponding normalized eigenstates are
\begin{equation}
    \begin{aligned}
    |\psi_n^{(+)}\rangle &= \cos\theta_n\, |e, n\rangle + \sin\theta_n\, |g, n+1\rangle, \\
    |\psi_n^{(-)}\rangle &= -\sin\theta_n\, |e, n\rangle + \cos\theta_n\, |g, n+1\rangle.
    \end{aligned}
\end{equation}
where the mixing angle \( \theta_n \) is defined by: \(\tan(2\theta_n) = \frac{2G\sqrt{n+1}}{\Delta}\). For the ground state $|g,0\rangle \text{ is the eigenstate with energy } E_0 = -\frac{1}{2} \hbar \omega_q$.

Suppose the initial state of the qubit-mechanical system in the \( n \)-excitation subspace is an arbitrary superposition:
\begin{equation}
|\psi(0)\rangle = \alpha |e,n\rangle + \beta |g,n+1\rangle.
\end{equation}
In terms of the eigenstates $|\psi_n^{(+)}\rangle$ and $|\psi_n^{(-)}\rangle$ the initial state can be written as 
\begin{equation}
    |\psi(0)\rangle = c_+ |\psi_n^{(+)}\rangle + c_- |\psi_n^{(-)}\rangle.
\end{equation}
where,
\begin{eqnarray}
    c_+ &=& \langle \psi_n^{(+)} | \psi(0) \rangle = \alpha \cos\theta_n + \beta \sin\theta_n, \\
    c_- &=& \langle \psi_n^{(-)} | \psi(0) \rangle = -\alpha \sin\theta_n + \beta \cos\theta_n.
\end{eqnarray}
Starting from this initial state, if we evolve the qubit-mechanical system under the Jaynes-Cummings Hamiltonian, then after some time t, the state becomes:
\begin{equation}
    |\psi(t)\rangle = e^{-iHt/\hbar} |\psi(0)\rangle = c_+ e^{-i \Omega_n t} |\psi_n^{(+)}\rangle + c_- e^{i \Omega_n t} |\psi_n^{(-)}\rangle.
\end{equation}

In the resonant case $\Delta =0$ and $n=0$,
\begin{equation}
    \begin{aligned}
    \lvert \psi_0^{(+)} \rangle &= \frac{1}{\sqrt{2}} \left( \lvert e,0 \rangle + \lvert g,1 \rangle \right), \\
    \lvert \psi_0^{(-)} \rangle &= \frac{1}{\sqrt{2}} \left( -\lvert e,0 \rangle + \lvert g,1 \rangle \right),
    \end{aligned}
\end{equation}

with eigenvalues:$E_0^{(\pm)} = \hbar \omega_b  \pm \hbar G.$ Decompose into the eigen basis:
\begin{equation}
\begin{aligned}
c_+ &= \langle \psi_0^{(+)} \lvert \psi(0) \rangle = \frac{1}{\sqrt{2}} (\alpha + \beta), \\
c_- &= \langle \psi_0^{(-)} \lvert \psi(0) \rangle = \frac{1}{\sqrt{2}} ( \beta-\alpha).
\end{aligned}
\end{equation}

For initial states \( \lvert \psi(0) \rangle = \lvert e,0 \rangle \), \(  \lvert g,0 \rangle \),  \( \lvert \psi_0^{(+)} \rangle \), and  \( \lvert \psi_0^{(-)} \rangle \) the state evolves to 

\begin{eqnarray}
\lvert \psi(t) \rangle &=& \cos(Gt) \lvert e,0 \rangle - i \sin(Gt) \lvert g,1 \rangle, \\
\lvert \psi(t) \rangle &=& \lvert g,0 \rangle, \quad \text{(stationary state)}, \\
\lvert \psi(t) \rangle &=& \frac{e^{-i G t}}{\sqrt{2}} \left( \lvert e,0 \rangle + \lvert g,1 \rangle \right)\quad \text{(stationary state)}, \\
\lvert \psi(t) \rangle &=&  \frac{e^{i G t}}{\sqrt{2}} \left( -\lvert e,0 \rangle + \lvert g,1 \rangle \right)\quad \text{(stationary state)}.
\end{eqnarray}

Below, we present some applications that follow from the above dynamics.


\begin{table*}[ht]
\footnotesize 
\setlength{\tabcolsep}{2pt}
\renewcommand{\arraystretch}{2.5}
\begin{tabular}{|c|c|c|c|c|c|c|c|}
\hline
\textbf{QUBIT} &
\makecell{\textbf{COUPLING}\\\textbf{TYPE}} &
\makecell{\textbf{MECHANICAL}\\\textbf{RESONATOR}}&
\makecell{\textbf{COUPLING}\\\textbf{ RATE}} &
$\boldsymbol{\omega_b/2\pi}$ &
$\boldsymbol{\omega_q/2\pi}$ &
$\mathbf{T_b}$ &
$\mathbf{T_{1q}}$
\\
\hline

\multirow{5}{*}{TRANSMON}
&
\multirow{4}{*}{\shortstack{CHARGE\\COUPLING}}
& MEMBRANE~\cite{Pirkkalainen2013}
& 9 MHz
& 71.8 MHz
& 3.9 GHz
& 77 $\mu$s
& 0.3 $\mu$s
\\ 

&
& SAW~\cite{doi:10.1126/science.aaw8415}
& 2.6 MHz
& 3.97 GHz
& 3.95 GHz
& 7.53 $\mu$s
& 22 $\mu$s
\\ 

&
& BAW~\cite{doi:10.1126/science.aao1511}
& 300 kHz
& 6.65 GHz
& 6.6 GHz
& 106 $\mu$s
& 27 $\mu$s
\\

&
& FLIP CHIP~\cite{Wollack2022}
& 10 MHz
& 2 GHz
& 2.26 GHz
& 1 $\mu$s
& 5 $\mu$s
\\ 
\hhline{~-------}

&
{\shortstack{PHASE\\COUPLING}}
& NANO BEAM~\cite{Bera2021}
& 15 kHz
& 6.6 MHz
& 6 GHz
& 167 ms
& 10 $\mu$s
\\
\hline

\multirow{2}{*}{FLUXONIUM}
&
\multirow{2}{*}{\shortstack{CHARGE\\COUPLING}}
& MEMBRANE~\cite{Gerashchenko2025}
& 1.5 kHz
& 4 MHz
& 2.35 GHz
& 6 ms
& 20 $\mu$s
\\ 

&
& FLIP CHIP~\cite{PRXLee}
& 14 MHz
& 700 MHz
& 800 MHz
& 1.5 $\mu$s
& 3.57 $\mu$s
\\
\hline
\end{tabular}
\vspace{.2cm}
    \caption{Some experimental realizations of transmon- and fluxonium-based hybrid electromechanical systems. Here, SAW denotes a surface acoustic wave resonator and BAW denotes a bulk acoustic wave resonator. The quantities $\omega_b$ and $T_b$ represent the resonance frequency and decay time of the mechanical resonator, respectively, while $\omega_q$ and $T_{1,q}$ denote the transition frequency and energy relaxation time of the qubit, respectively.}
    \label{fig:table}
\end{table*}

\subsubsection{State transfer and generation of qubit-mechanical entangled states}

\begin{figure}[htbp]
    \centering
    \includegraphics[width=40mm]{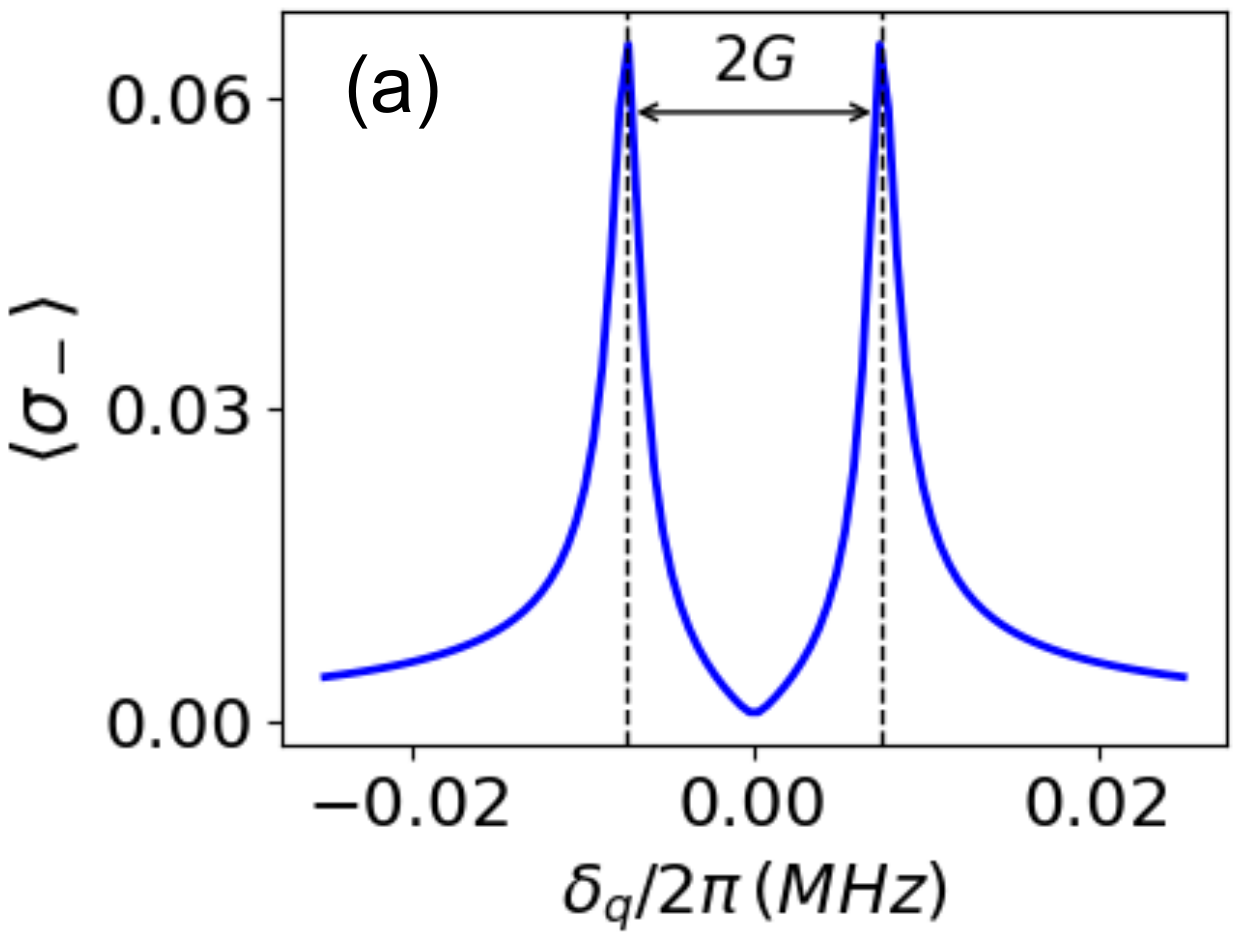}
    \vspace{-10mm}
    \includegraphics[width=40mm]{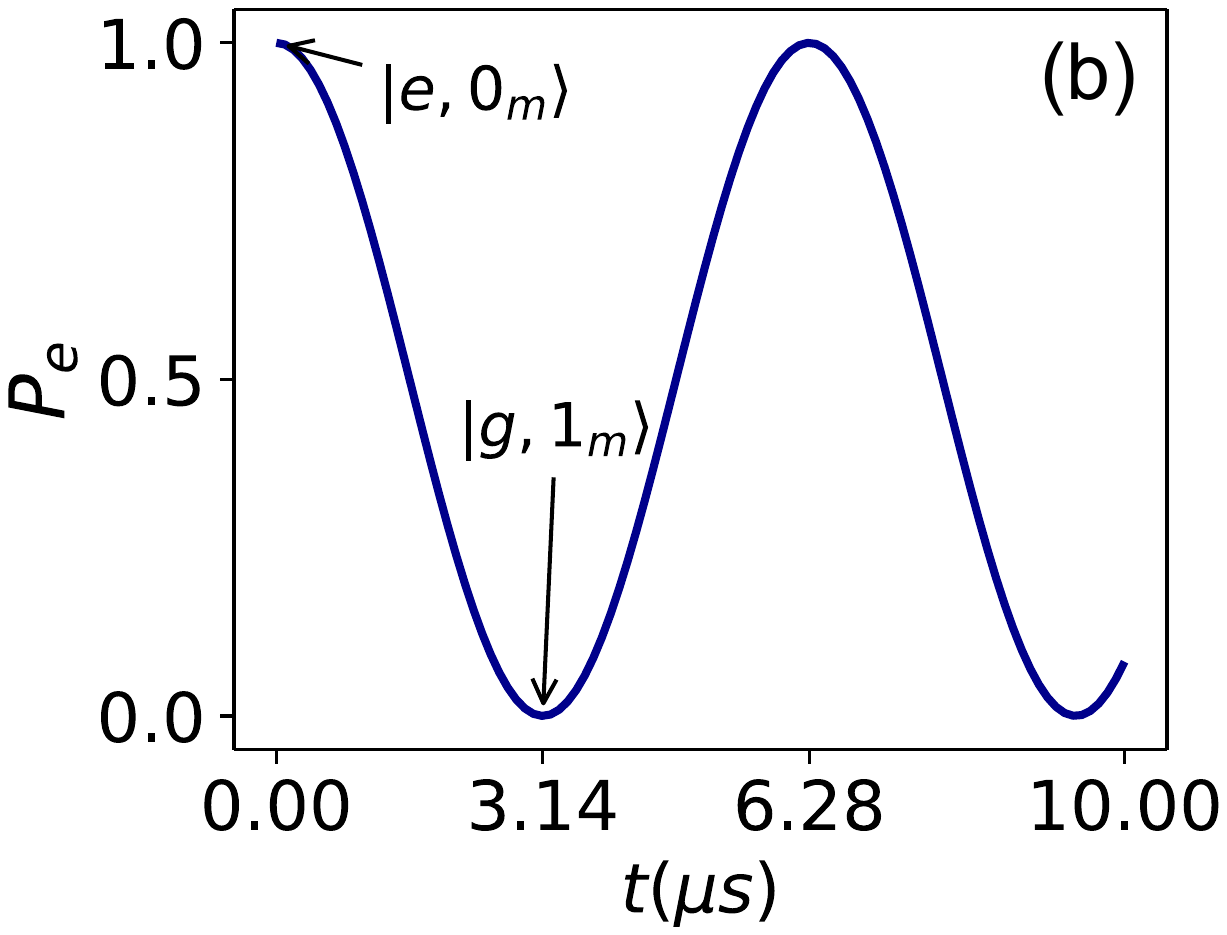}
    \vspace{-10mm}
    \caption{
    (a) Mode splitting of the qubit–mechanical system in the semi-classical treatment.  
    (b) Evolution of the excited state population $P_e$, showing the coherent transfer of excitations between the qubit and the resonator in the quantum regime.  
    Parameters used:  
    (a) $\delta_q=\omega_d-\omega_q$, $\omega_q = \omega_b = 4.5~\text{GHz}$, $G = 7.5~\text{MHz}$, $\gamma_m = \gamma = 1~\text{MHz}$, and $\epsilon = 0.1~\text{MHz}$.  
    (b) $G = 0.5~\text{MHz}$.
    }
    \label{fig:Mode_split}
\end{figure}

When we prepare the qubit-mechanical resonator system in the ground state \( \lvert g,0 \rangle \) or in one of the eigenstates of the Jaynes-Cummings interaction, \( \lvert \psi_0^{(+)} \rangle \) or \( \lvert \psi_0^{(-)} \rangle \), the system remains in the same state up to a global phase factor during its evolution. 
These states are known as stationary states, and the expectation value of any observable in such states does not change with time. In other words, measuring an observable in the stationary basis yields time-independent results.
However, if the system is initialized in the state \( \lvert e,0 \rangle \) or equivalently \( \lvert g,1 \rangle \), the state undergoes coherent oscillations between \( \lvert e,0 \rangle \) and \( \lvert g,1 \rangle \). 
These vacuum Rabi oscillations, as shown in Fig .\ref{fig:Mode_split}(b), can be exploited as a resource to swap quantum states between the qubit and the resonator, or to generate entanglement between them. This is achieved by allowing the qubit to interact with the mechanical resonator for a precise duration.
For example, at time \( t = \pi / (2G) \), the excitation is fully transferred: the qubit transitions from the excited to the ground state, while the resonator becomes excited. 
Similarly, at time \( t = \pi / (4G) \), the system evolves into a maximally entangled state of the qubit and the resonator.

Experimentally, one can observe Rabi oscillations of the qubit through its spectrum which is obtained from the excited-state population,

\begin{equation}
P_e = \frac{1}{2} \left( \langle \hat{\sigma}_z \rangle + 1 \right)
\end{equation}

\begin{figure}[!hbt]
    \centering
    \includegraphics[width=60mm]{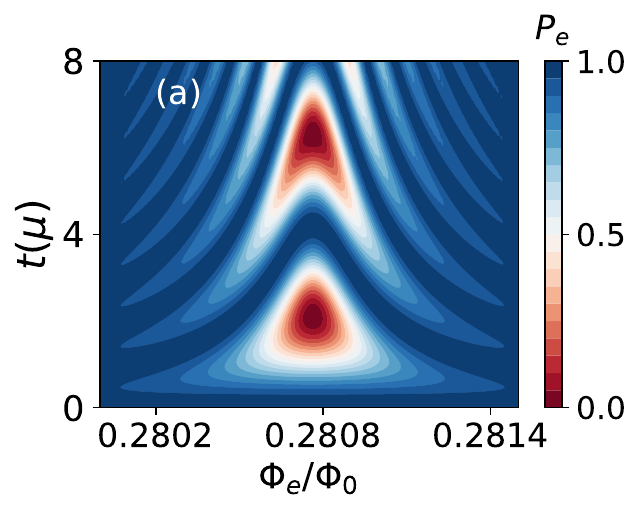}
    \includegraphics[width=60mm]{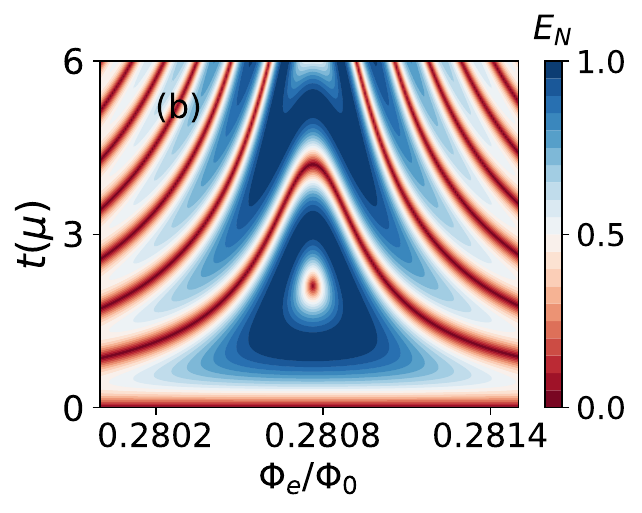}
    \caption{
    (a) Rabi oscillation between the qubit and mechanical resonator for different applied flux values. $P_e$ is the probability of the qubit being in the excited state.  
    (b) The corresponding entanglement measure $E_n$ for different applied flux values.  
    Parameters used: $\omega_b = 4.5~\text{GHz}$, $g_T = 1~\text{MHz}$, $\Omega_R = 10~\text{kHz}$, $E_{J\text{max}} = 10~\text{GHz}$, $E_C = 0.5~\text{GHz}$, $d=0$. Adapted from \cite{Nongthombam2025}.
    }
    
    \label{fig:Rabi}
\end{figure}

In Fig.\ref{fig:Rabi}(a), the Rabi oscillations for the transmon–mechanical system are shown.
As expected, when we vary the qubit frequency by tuning the external flux, the oscillations diminish. 
This occurs because the resonance condition between the qubit and the resonator is reduced. The entanglement measure, defined as $E_N(\hat{\rho}) = \log_2[2N(\rho) + 1]$, where \( N(\rho) \) represents the absolute value of the sum of the negative eigenvalues of the partial transpose of the joint density matrix \cite{vidal2002computable, plenio2005introduction, plenio2005logarithmic}, of the qubit-mechanical state during the Rabi oscillation is also shown in Fig.\ref{fig:Rabi}(b). 

In the presence of a thermal environment, where the system is coupled to a finite-temperature cold bath, the Rabi oscillations will eventually decay. 
To model this decay, we evolve the Jaynes–Cummings interaction Hamiltonian (with an added drive term) using the Lindblad master equation. The dynamics of the transmon–mechanical system are governed by,

\begin{align}
\frac{d\hat{\rho}_i}{dt} = -\frac{i}{\hbar} [\hat{H}_{j}, \hat{\rho}_i] + \gamma_i\,\mathcal{D}[\lvert g_i\rangle\langle e_i\rvert]\,\hat{\rho}_i + \gamma_{i\phi}\,\mathcal{D}[\lvert e_i\rangle\langle e_i\rvert]\,\hat{\rho}_i \notag \\+ \gamma_m(n_{\text{th}} + 1)\,\mathcal{D}[\hat{b}]\,\hat{\rho}_i + \gamma_m n_{\text{th}}\,\mathcal{D}[\hat{b}^\dagger]\,\hat{\rho}_i,
\end{align}

where \( i \in \{\text{T}, \text{F} \} \) corresponds to the transmon or fluxonium qubit, and \( j \in \{\text{TM}, \text{FM} \} \) denotes the corresponding system Hamiltonian. The Lindblad dissipator is defined as

\begin{equation}
\mathcal{D}[\hat{L}]\,\hat{\rho} = \hat{L} \hat{\rho} \hat{L}^\dagger - \frac{1}{2} \left( \hat{L}^\dagger \hat{L} \hat{\rho} + \hat{\rho} \hat{L}^\dagger \hat{L} \right).
\end{equation}

Here, \( \gamma_T \) (\( \gamma_F \)) is the relaxation rate for the transmon (fluxonium) qubit, \( \gamma_\phi \) is the pure dephasing rate, \( \gamma_m \) is the mechanical damping rate, and \( n_{\text{th}} \) is the thermal occupation number of the mechanical mode. 
The thermal occupation of the qubit is neglected under the assumption that only its GHz transition frequency is considered.
The Rabi oscillation decays over time. To enable meaningful quantum operations in the qubit–mechanical hybrid system, the coherence time must be long enough to sustain several oscillation cycles.

Instead of observing the coherent exchange of excitations between the qubit and the mechanical resonator at the single-phonon level, one can also observe mode splitting between the qubit and the resonator in the semiclassical regime. 
In this regime, the qubit sees the resonator as a classical drive with amplitude \( \langle b \rangle \). The interaction term in the Jaynes–Cummings Hamiltonian can be equivalently written as  
\(
\hbar G ( \sigma_+ \langle b \rangle + \sigma_- \langle b^\dagger \rangle ),
\)
which corresponds to a qubit driven by the classical resonator field.
To probe the mode splitting, a weak probe drive is applied via the term  
\(
H_d = \Omega_R \cos(\omega_d t)\, \sigma_x,
\). 
The dynamics of the coupled qubit-resonator system under the semiclassical approximation and in the rotating frame of the drive frequency \( \omega_d \) are given by:

\begin{align}
    \frac{d}{dt} \braket{b} &= -\left( \frac{\gamma_m}{2} + i(\omega_b - \omega_d) \right) \braket{b} - i G \braket{\sigma_-}, \\
    \frac{d}{dt} \braket{\sigma_-} &= -\left( \frac{\gamma + \gamma_\phi }{2} + i(\omega_q - \omega_d) \right) \braket{\sigma_-} + i G \braket{b} \braket{\sigma_z}+ i \Omega_R \braket{\sigma_z}, \\
    \frac{d}{dt} \braket{\sigma_z} &= -\gamma (\braket{\sigma_z} + 1) - 2i G \braket{b} \braket{\sigma_+}+ 2i G \braket{b^*} \braket{\sigma_-} \notag \\ 
    &+ 2i \Omega_R \left( \braket{\sigma_-} - \braket{\sigma_+} \right).
\end{align}

Here, \( \epsilon \) is the strength of the drive, and \( \gamma_m \), \( \gamma \), and \( \gamma_\phi \) are the mechanical damping, qubit relaxation, and qubit dephasing rates, respectively. 
The excited-state population is plotted in Fig.~\ref{fig:Mode_split}, showing a mode splitting of \( 2G \).
\subsubsection{Probing the phonon number distribution of the mechanical resonator}

In the \textit{dispersive regime}, where the detuning between the qubit and mechanical resonator frequencies is much larger than the resonant coupling strength, the phonon number distribution of the mechanical resonator can be measured by probing the qubit's excited state probability spectrum, \( P_e \). 
Similar to the case of observing Rabi oscillations, we drive the qubit with a microwave drive at frequency \( \omega_s \) and amplitude \( \Omega_R \). Simultaneously, the mechanical resonator is driven resonantly with a pulse at frequency \( \omega_d \) and amplitude \( \epsilon \).
For a transmon qubit, using a similar relation as in Eq.~\eqref{dispersive}, the Hamiltonian in the drive frame and under the rotating wave approximation (RWA) is given by:
\begin{equation}
    \label{eqn:dispersive_QM}
    \hat{H}_d = \frac{\hbar}{2}(\Delta_T + 2\chi\, \hat{b}^\dagger \hat{b})\, \hat{\sigma}_z + \hbar \Delta_m\, \hat{b}^\dagger \hat{b} + \hbar \epsilon (\hat{b}^\dagger + \hat{b}) + \hbar \frac{\Omega_R}{2} \hat{\sigma}_x,
\end{equation}
where
\begin{equation}
    \label{eqn:trans_dis}
    \chi = -\frac{G^2 E_C/\hbar}{\delta(\delta - E_C/\hbar)}, \quad
    \Delta_m = \omega_b - \omega_d + \frac{G^2}{\delta - E_C/\hbar}.
\end{equation}
Here,
\[
\Delta_T = \omega_{0T} + \frac{G^2}{\delta} - \omega_s, \quad \delta = \omega_{0T} - \omega_b.
\]
The coupling strength \( G \) between the qubit and the resonator is fixed at a chosen flux bias point.
The last two terms in Eq.~\eqref{eqn:dispersive_QM} correspond to the mechanical phonon pump and qubit spectroscopy, respectively.

\begin{figure}[!hbt]
    \centering
    \includegraphics[width=40mm]{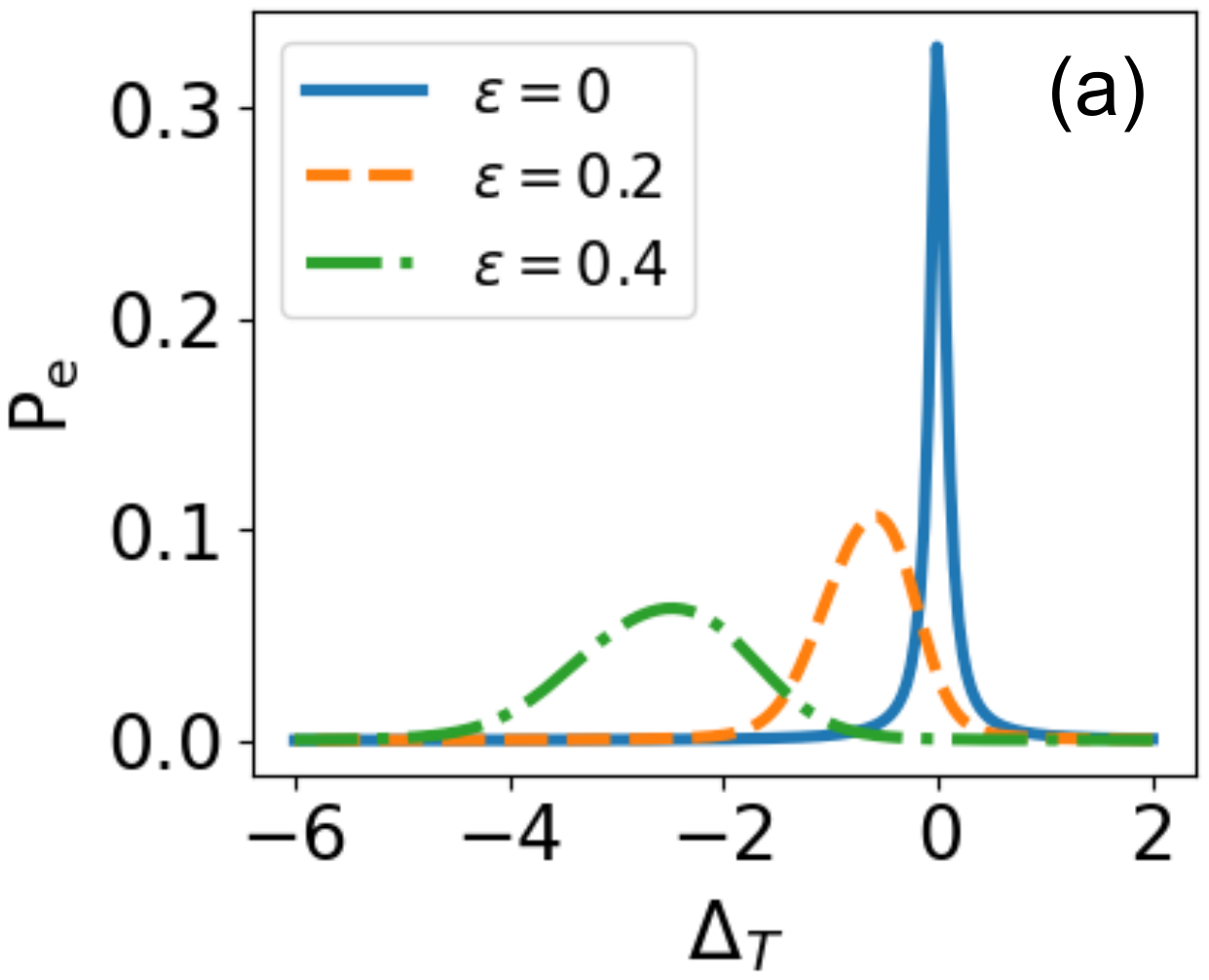}
        \vspace{-8mm}
    \includegraphics[width=40mm]{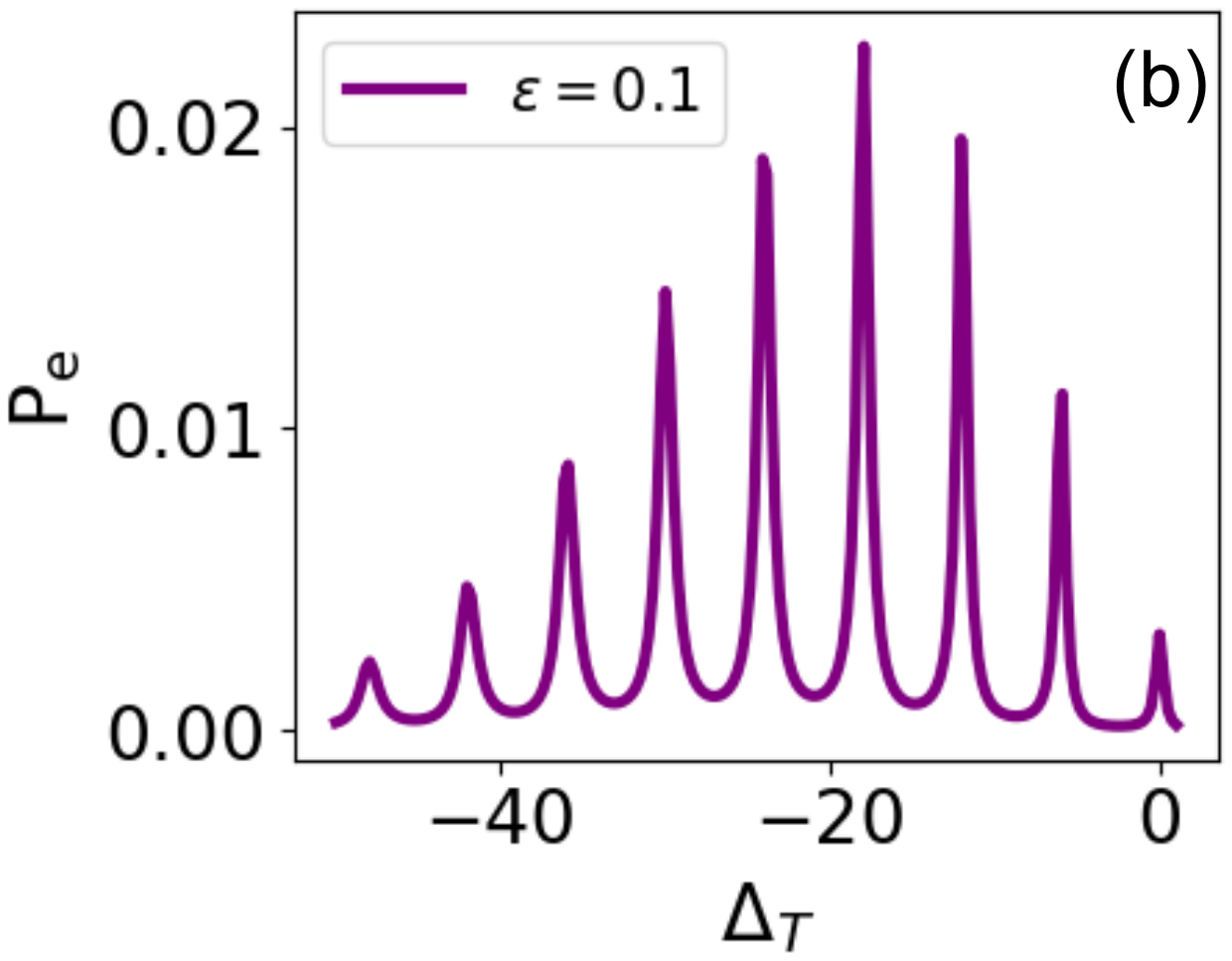}
        \vspace{-8mm}
    \caption{
    Excited-state population of the qubit as a function of the detuning between the qubit and the drive for (a) weak and (b) strong coupling.  
    Parameters used:  
    (a) $\chi = 0.1~\text{MHz}$, $\Delta_T = 0$, $\Omega_R = \gamma_b = \gamma = 0.1~\text{MHz}$.  
    (b) $\chi = 3~\text{MHz}$, $\Delta_T = -\chi$, $\Omega_R = \gamma_b = \gamma = 0.1~\text{MHz}$, $\epsilon = 0.1$. Adapted from \cite{cqed1}
    }.
    \label{fig:number_split}
\end{figure}

The qubit spectrum, which reveals the phonon number distribution, is shown in Fig.~\ref{fig:number_split}. 
For weak dispersive coupling (\( \chi < \gamma \)), the frequency shift in the qubit is governed by the average phonon occupancy \( 2\chi \langle b^\dagger b \rangle \). 
In the strong dispersive coupling regime (\( \chi > \gamma \)), individual Fock states are spectrally resolved.
As the average phonon number increases, linewidth broadening in the qubit spectrum also increases. 
Notably, the spectral lineshape transitions from Lorentzian to Gaussian, with linewidth scaling from \( \bar{m} \) to \( \sqrt{\bar{m}} \). 
A similar phonon-number-dependent shift of the qubit frequency can be observed in the fluxonium qubit. 
In that case, the Lamb and AC Stark shifts receive significant contributions from higher energy levels, which must be evaluated numerically.
Phonon excitations inferred from the qubit spectrum provide a useful resource for quantum metrology, where interactions with external forces are encoded in these excitations.

\subsection{Coupling through the phase degree of freedom}

\begin{figure}[!hbt]
    \centering
    \includegraphics[width=90mm]{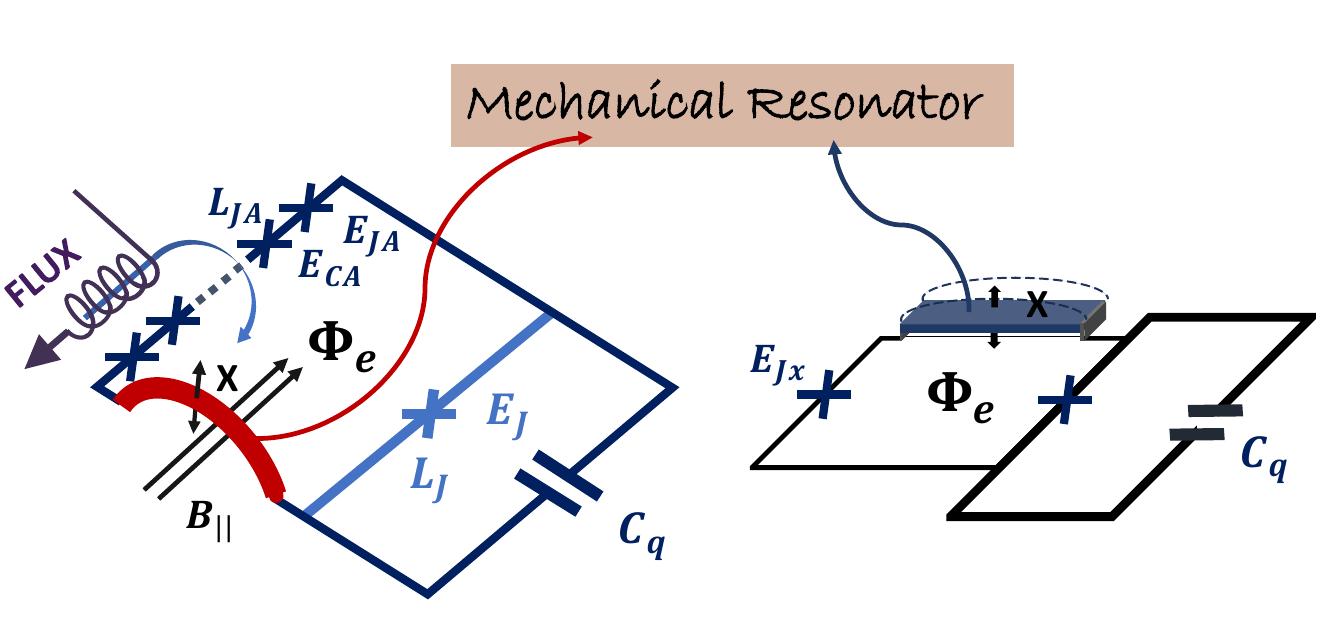}
    \caption{
    Hybrid electromechanical system realized through coupling via the phase degree of freedom.  
    (left) Implementation with fluxonium, and  
    (right) implementation with transmon.  
    The suspended mechanical beam, in the presence of a magnetic field, induces a motion-dependent flux, leading to coupling between the mechanical resonator and the qubit. Adapted from \cite{Nongthombam2025}.
    }
    \label{fig:system_qubit_mech_phase}
\end{figure}

When an external magnetic flux is applied, the total phase accumulated around the superconducting loop, such as the SQUID loop of a flux-tunable transmon or the loop in a fluxonium qubit, must satisfy the flux quantization condition. According to this rule, the sum of the phase differences across the Josephson junction(s) and the total magnetic flux threading the loop must equal an integer multiple of \(2\pi\).
Suppose a small segment of the loop is mechanically suspended, forming a resonator that can vibrate. 
If a static magnetic field is applied in the loop, then the motion of this suspended segment induces an additional motional flux due to its displacement. 
This motional flux adds to the externally applied flux (typically introduced via a local flux line), and both contribute to the total flux threading in the loop.
The flux quantization condition then requires that the sum of the phase differences across the junctions and the total flux, including both the external and motional contributions, continue to satisfy the quantization rule. 
This dependence introduces a coupling between the mechanical resonator and the superconducting phase (and hence the qubit), mediated by the flux degree of freedom \cite{PhysRevResearch.2.023335}.
When this flux dependence is incorporated into the qubit Hamiltonian (for either the transmon or fluxonium), it gives rise to an interaction term between the mechanical resonator and the qubit, enabling coherent coupling via the phase degree of freedom. The setup for the phase degree of freedom–induced electromechanical system is shown in Fig.~\ref{fig:system_qubit_mech_phase}.
In the case of a flux-tunable transmon qubit, the Hamiltonian incorporating the momotion-induced flux is given by \cite{transmon1}, 
\begin{equation}
    \label{phase_transmon_mech_coupling}
    \hat{H}_\text{Tm} = 4 E^\text{T}_C \,(\hat{n}^\text{T})^2 - E_J^{\text{max}} \cos\left( \frac{\pi(\Phi_e + \hat\Phi_b)}{\Phi_0} \right) \cos\hat{\theta}_\text{T}.
\end{equation}
A symmetric transmon is considered where the two Josephson junctions have equal Josephson energies. 
The momotion-induced flux is given by $\hat\Phi_b = \beta_0 B l \hat{x}$, where $\beta_0$ is a geometric factor, \( B \) is the applied magnetic field, \( l \) is the length of the suspended resonator, and \( \hat{x} \) is its displacement.
Assuming the contribution from the momotion-induced flux is small compared to the applied flux, we expand the Hamiltonian Eq. \eqref{phase_transmon_mech_coupling} to first order in \( \hat\Phi_b \), yielding:
\begin{align}
    \label{phase_transmon_mech_coupling_approx1}
    \hat{H}_\text{Tm} \approx 4 E^\text{T}_C \,(\hat{n}^\text{T})^2 - E_J^{\text{max}} \cos\left( \frac{\pi\Phi_e }{\Phi_0} \right) \cos\hat{\theta}_\text{T}\notag \\
    - \alpha \sin\left( \frac{\pi\Phi_e }{\Phi_0} \right) x \cos\hat{\theta}_\text{T} + \hbar \omega_b \hat{b}^\dagger \hat{b},
\end{align}
where $\alpha = \pi E_J^{\text{max}} \beta_0 B l / \Phi_0$ and $x = x_0 (b^\dagger + b)$, with $x_0 = \sqrt{\hbar / (2 m \omega_b)}$ being the zero-point fluctuation amplitude of the mechanical resonator.
The first two terms in Eq.~\eqref{phase_transmon_mech_coupling_approx1} describe the standard transmon qubit. Rewriting the Hamiltonian in the eigen basis \( \{ |g_T\rangle, |e_T\rangle \} \) of the unperturbed transmon, we obtain:
\begin{align}
    \label{phase_transmon_mech_coupling_eigenform}
    \hat{H}_{\text{Tm}} &= \hbar \omega_b \hat{b}^\dagger \hat{b} + \sum_{i = g_T, e_T} E_i^{\text{T}} |i_T\rangle\langle i_T| \notag \\ 
    &  + \hbar g_{\text{Tm}} \sum_{i,j = g_T, e_T} \theta_{ij}^{\text{Fm}} |i_T\rangle\langle j_T| (\hat{b} + \hat{b}^\dagger),
\end{align}
where \( \theta_{ij}^{\text{Fm}} = \langle i_T| \cos\theta | j_T\rangle \), and the coupling strength is given by
\[
    g_{\text{Tm}} = \frac{\alpha ~x_0}{\hbar}\sin\left( \frac{\pi\Phi_e }{\Phi_0} \right) .
\]
This resembles charge-based coupling, but here the coupling matrix elements arise from the phase degree of freedom.  
The matrix elements $\theta_{ij}^{\text{Tm}}$ at the external flux $\Phi_e = 0.3\,\Phi_0$ differ from those in charge-mediated coupling, as only the diagonal elements contribute significantly, resulting in a longitudinal interaction.
By retaining only the non zero longitudinal terms in Eq.~\eqref{phase_transmon_mech_coupling_eigenform} and expressing the result in terms of Pauli operators, we arrive at:
\begin{equation}
\label{H_Tm}
    \hat{H}_\text{Tm} = \hbar \omega_b \hat{b}^\dagger \hat{b} + \frac{\hbar \omega_{0T}}{2} \hat{\sigma}_z + \hbar G_\text{Tm} \, \hat{\sigma}_z (\hat{b}^\dagger + \hat{b}),
\end{equation}
where \( \omega_{0T} = (E_e^{\text{T}} - E_g^{\text{T}})/\hbar \), and the effective coupling strength is 
\[
    G_\text{Tm} = \frac{1}{2}(\theta_{ee}^{\text{Tm}} - \theta_{gg}^{\text{Tm}}) \, g_{\text{Tm}}.
\]

Similar to the transmon-mechanical coupling analysis, a fluxonium qubit coupled to a suspended mechanical resonator can be treated analogously. The Hamiltonian of the fluxonium in the presence of a motionally induced flux can be written as
\begin{equation}
\label{Fluxonium-mech1}
\hat{H}_\text{Fm} = 4E^\text{F}_C (\hat{n}^\text{F})^2 - E_J \cos(\hat{\theta}^\text{F}) + \frac{1}{2} E_L \left(\hat{\theta}^\text{F} - \frac{2\pi(\Phi_e + \hat\Phi_b)}{\Phi_0}\right)^2,
\end{equation}
where \( E_C = \frac{e^2}{2C_q} \), \( E_J = \frac{\Phi_0^2}{L_J} \),  
\( E_L = \frac{\Phi_0^2}{L} \), and \( \Phi_0 = \frac{h}{2e} \).

Note that, since the displacement \( \hat{x}(t) \) of the resonator is time-dependent, it is convenient to place the external flux term in the quadratic potential, while \( \Phi_e \) is retained in the cosine term \cite{PhysRevLett.129.037205, Nongthombam2025, Xinyuan_You_2019}.  
When \( \hat\Phi_b(t) \) is added to the cosine term in Eq.~\eqref{Fluxonium-mech1}, a time derivative \( \dot{\hat\Phi}_b(t) \) appears in the Hamiltonian. 
This can be eliminated by imposing an irrotational constraint on the flux variable. Equivalently, this means placing the time-dependent flux term in the quadratic potential rather than the cosine potential.
Expanding Eq.~\eqref{Fluxonium-mech1}, we obtain:
\begin{align}
\label{Fluxonium-mech2}
\hat{H}_\text{Fm} &= 4E^\text{F}_C (\hat{n}^\text{F})^2 - E_J \cos\left(\hat{\theta}^\text{F} + \frac{2\pi\Phi_e}{\Phi_0} \right)\notag \\ 
    &
+ \frac{E_L}{2} (\hat{\theta}^\text{F})^2 + \hbar g_{\Phi} (\hat{b} + \hat{b}^\dagger) \hat{\theta}^F + \hbar \omega_b \hat{b}^\dagger \hat{b}.
\end{align}
The single-phonon transverse coupling strength between the fluxonium and the resonator is given by  
\begin{equation}
g_{\Phi} = \frac{2\pi ~E_L B l x_0}{\hbar~\Phi_0},
\end{equation}
where \( x_0 = \sqrt{\frac{\hbar}{2m\omega_b}} \) is the quantum zero-point displacement of the resonator.
The first three terms in Eq.~\eqref{Fluxonium-mech2} describe the standard fluxonium qubit. Rewriting the Hamiltonian in the eigen basis \( \{ |g_F\rangle, |e_F\rangle \} \) of the unperturbed fluxonium, we obtain:
\begin{align}
\label{phase_fluxonium_mech_coupling_eigenform}
\hat{H}_\text{Fm} &= \hbar \omega_b \hat{b}^\dagger \hat{b} + \sum_{i = g_F, e_F} E_i^{\text{F}} |i_F\rangle\langle i_F| \notag \\&\quad
+ \hbar g_{\text{Fm}} \sum_{i,j = g_F, e_F} \theta_{ij}^{\text{Fm}} |i_F\rangle\langle j_F| (\hat{b} + \hat{b}^\dagger),
\end{align}
where \( \theta_{ij}^{\text{Fm}} = \langle i_F| \theta^F | j_F\rangle \).
The matrix elements at $\Phi_e = 0.3\,\Phi_0$ are dominated by the diagonal elements, resulting in longitudinal coupling, whereas at $\Phi_e = 0.5\,\Phi_0$, the off-diagonal elements dominate, leading to transverse coupling.  
At these two flux points, the effective Hamiltonians for the coupled system take the form:
\begin{equation}
\hat{H}_\text{FmL} = \hbar \omega_b \hat{b}^\dagger \hat{b} + \frac{\hbar \omega_{0F}}{2} \hat{\sigma}_z + \hbar G_\text{FmL} \, \hat{\sigma}_z (\hat{b}^\dagger + \hat{b}),
\label{Eq:Long}
\end{equation}
and
\begin{equation}
\hat{H}_\text{Fmt} = \hbar \omega_b \hat{b}^\dagger \hat{b} + \frac{\hbar \omega_{0F}}{2} \hat{\sigma}_z + \hbar G_\text{Fmt} \, \hat{\sigma}_x (\hat{b}^\dagger + \hat{b}).
\end{equation}
The corresponding coupling constants are given by
\[
G_{\text{FmL}} = \frac{g_{\Phi}}{2} (\theta_{ee}^{\text{Fm}} - \theta_{gg}^{\text{Fm}})
\]
and
\[
G_{\text{Fmt}} = g_\Phi \theta_{eg}^{\text{Fm}},
\]
where \( \theta_{eg}^{\text{Fm}} = \theta_{ge}^{\text{Fm}} \).
In both the transmon and fluxonium cases, the coupling to the mechanical resonator at an externally applied flux \( \Phi_e = 0.3\,\Phi_0 \) is purely longitudinal.

Some of the experimental realizations of transmon- and fluxonium-based systems coupled to mechanical resonators via the phase degree of freedom are shown in Fig.~\ref{fig:table}. While a transmon coupled to a suspended nanomechanical beam through the phase degree of freedom has been experimentally demonstrated, the corresponding fluxonium-based implementation has, to the best of our knowledge, not yet been realized experimentally. Several experimental challenges may contribute to this difficulty. First, the control and readout of fluxonium qubits operating at MHz transition frequencies are experimentally demanding, particularly in 3D cavity architectures. Second, a strong and precisely aligned magnetic field is required to bias the fluxonium without introducing vortices, additional losses, or Josephson-junction degradation. Third, precise control of the magnetic flux is necessary to access and observe the avoided crossings. Finally, the mechanical resonator must be integrated into the fluxonium loop while preserving both qubit coherence and mechanical performance, which  introduces significant fabrication and design challenges.

Below, we discuss several applications of the longitudinal interaction.

\subsubsection{Ground-state cooling of the mechanical resonator}
The type of interaction shown in Eq.~\eqref{Eq:Long} has several important applications. One such application is ground-state cooling of the mechanical resonator \cite{jaehne2008ground}. 
The suspended mechanical resonator considered here typically has a frequency in the few MHz range. 
At  millikelvin temperatures, as achieved in a typical dilution refrigerator, the resonator remains in a thermal state at steady state due to its low energy scale. In contrast, both the fluxonium and transmon qubits have transition frequencies in the GHz regime at this flux bias point and therefore remain close to their ground states under thermal equilibrium. 
As a result, the qubit can effectively act as a cold reservoir for the mechanical resonator. 
This setup enables cooling of the resonator via the longitudinal interaction. 
Cooling is achieved by applying a red-detuned drive to the qubit. Due to this detuning, the qubit is excited to a higher energy level by absorbing incoming photons from the external drive as well as phonons from the mechanical oscillator. As a result, the mechanical resonator loses phonons to the qubit, leading to cooling. 
A detailed analysis of this cooling process can be found in \cite{PhysRevA.104.023509}. 

\subsubsection{Encoding a qubit state into a mechanical coherent state}
Since the interaction term in Eq.~\eqref{Eq:Long} commutes with the bare qubit part of the Hamiltonian, it can be utilized to encode the qubit state into the mechanical resonator. 
This is achieved by modulating the qubit-resonator interaction. Under the rotating wave approximation (RWA), such modulation allows us to describe the system in the interaction frame.
In the transmon case, this modulation can be straightforwardly implemented by applying a weak parametric AC bias of the form \( \Phi_e = \Phi_{\text{ac}} \cos(\omega_{\text{ac}} t) \), where \( \pi \Phi_{\text{ac}} / \Phi_0 \ll 1 \). 
This results in a time-dependent coupling strength that can be written as
\begin{equation}
G_\text{Tm}(t) = \frac{\pi \alpha x_0 \, \Phi_{\text{ac}}}{2\Phi_0} \cos(\omega_{\text{ac}} t) \left( \theta_{ee}^{\text{Tm}} - \theta_{gg}^{\text{Tm}} \right).
\end{equation}
Substituting this modulated coupling into the Hamiltonian Eq. ~\eqref{H_Tm}, and transforming to the interaction frame defined by the AC drive \( U = e^{i \omega_{ac} t b^\dagger b} \) and the qubit frame, we obtain
\begin{equation}
\label{H^I_Tm_t}
\hat{H}^I_{\text{Tm}} = \hbar G_0 \text{cos}(\omega_{ac}t)  \, \hat\sigma_z (\hat{b} e^{-i\omega_bt} + \hat{b}^\dagger e^{i\omega_bt}),
\end{equation}
where the effective coupling constant is
\begin{equation}
G_0 = \frac{\pi \Phi_{\text{ac}}\alpha x_0}{2\Phi_0} \left( \theta_{ee}^{\text{Tm}} - \theta_{gg}^{\text{Tm}} \right).
\end{equation}
Setting the modulation frequency to resonance $\omega_{ac}=\omega_b$ and neglecting the fast-rotating terms under the condition \( 2\omega_b \gg G_0 \), we obtain
\begin{equation}
\label{H^I_Tm}
\hat{H}^I_{\text{Tm}} = \hbar G_0 \, \hat\sigma_z (\hat{b}  + \hat{b}^\dagger ).
\end{equation}
The system evolves as $|\Psi_{Tm}(t)\rangle=U(t)|\Psi\rangle_0$, where $U(t)=exp[-iG_0t\sigma_z(b^\dagger+b)]$ and $|\Psi\rangle_0=|g\rangle|0_b\rangle$ is the initial state of the system.
If the qubit is initially in the ground state \( |g\rangle \) and the mechanical resonator is in its vacuum state \( |0_b\rangle \), then after a time \( t \), the qubit remains in the ground state while the resonator evolves into a coherent state \( |\beta_b = i G_0 t \rangle \). 
Similarly, if the qubit starts in the excited state \( |e\rangle \), the resonator evolves into a coherent state with opposite amplitude \( |\beta_b = -i G_0  t \rangle \).

\begin{center}
    \( |g, 0_b \rangle_0 \longrightarrow |g, \beta_b = i G_0  t \rangle_t \)\\[0.4cm]
    \( |e, 0_b \rangle_0 \longrightarrow |e, \beta_b = -i G_0  t \rangle_t \)
\end{center}

\begin{figure*}[htbp]
    \centering
    \includegraphics[width=60mm]{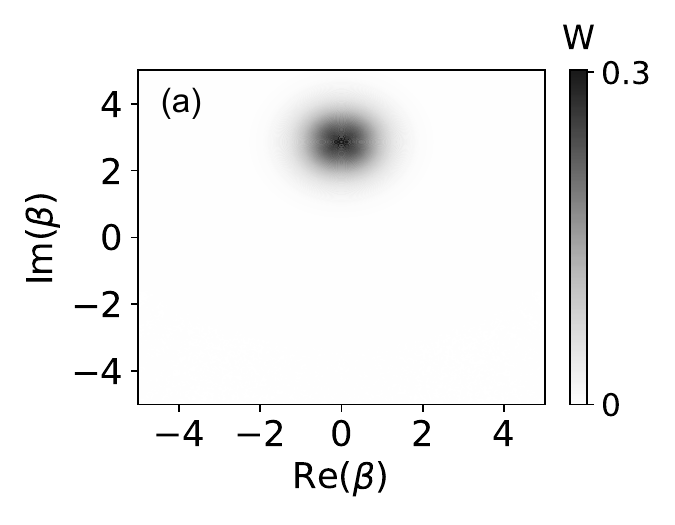}
    \includegraphics[width=60mm]{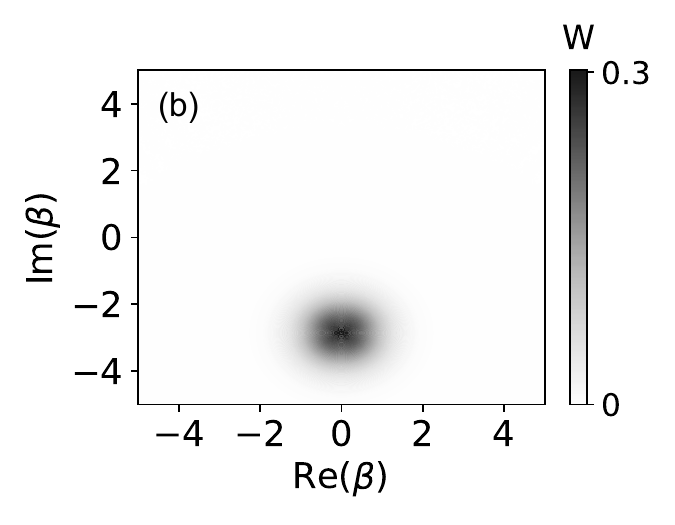}
    \includegraphics[width=60mm]{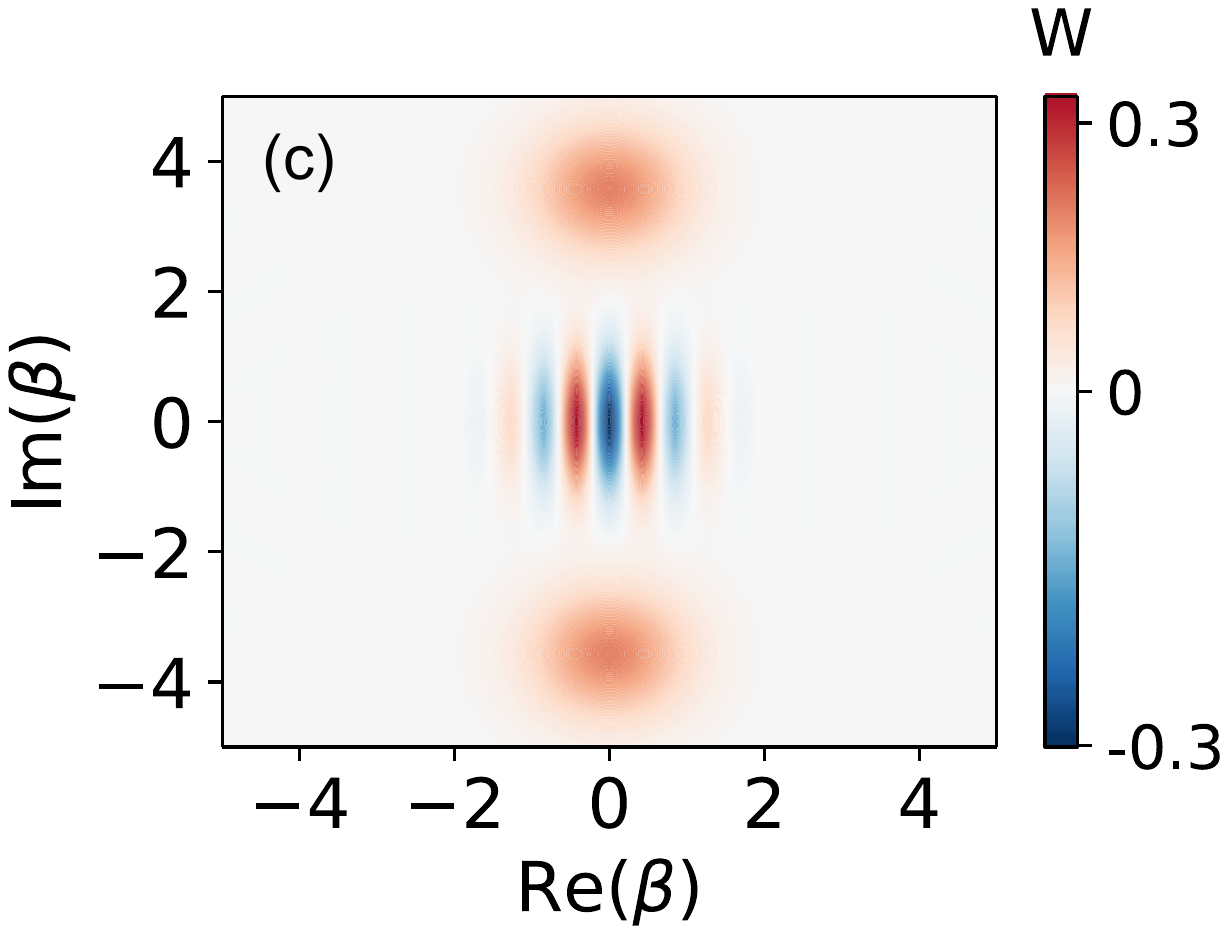}
    \includegraphics[width=60mm]{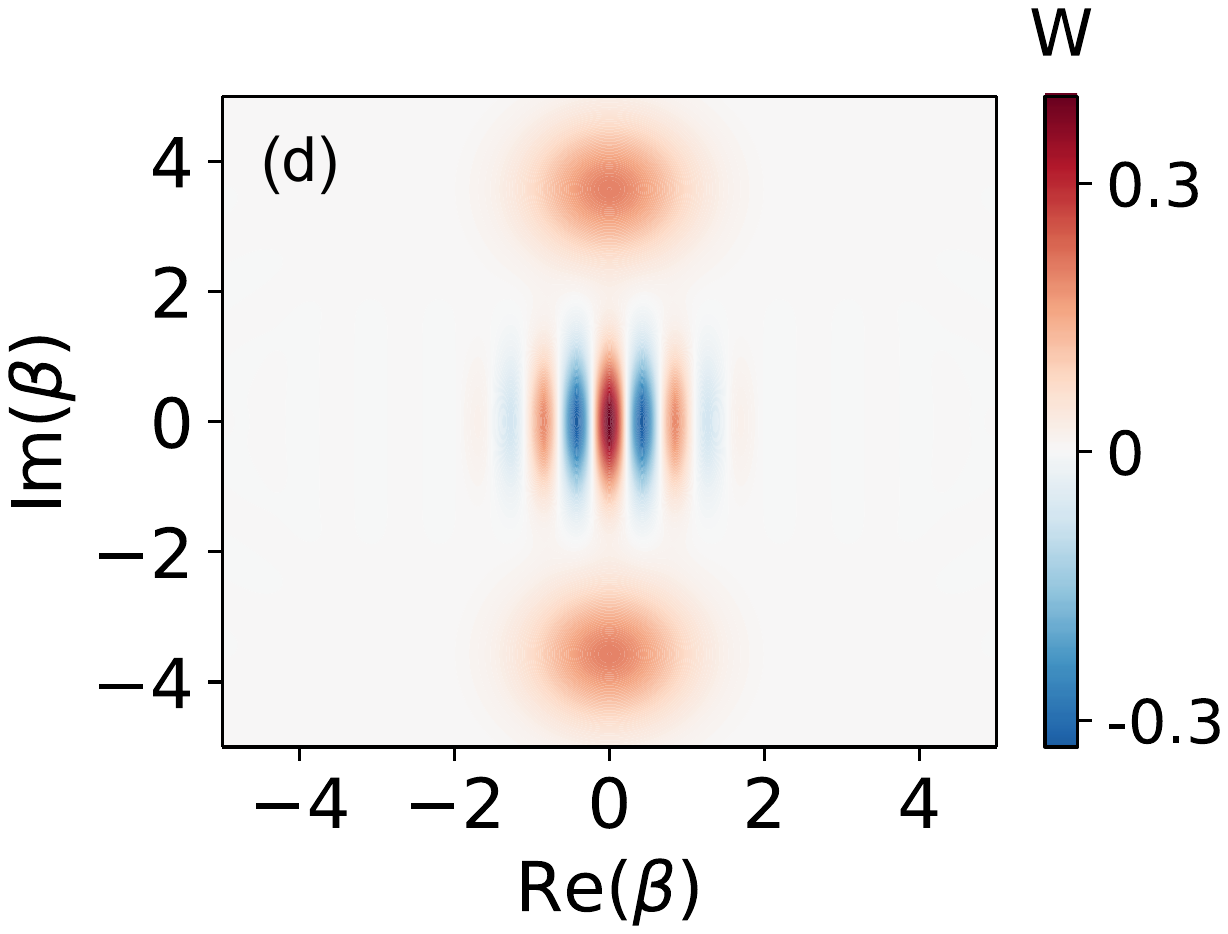}
    \vspace{-30mm}
    \caption{Coherent displacement of the mechanical resonator when the qubit is in the (a) ground state and (b) excited state. Panels (c) and (d) show the even and odd cat states of the resonator, respectively, projected after measurement of the qubit in the ground and excited states.} 
    \label{fig:coherent_mech}
\end{figure*}

Therefore, we observe that the ground and excited states of the qubit become associated with coherent states of the resonator that have equal magnitude but opposite phase as shown in Fig. \ref{fig:coherent_mech} (a-b). This effectively encodes the qubit state into the mechanical resonator state.
This protocol requires additional ground-state cooling since, even at dilution refrigerator temperatures ($\sim 15~\mathrm{mK}$), the thermal phonon occupation is still on the order of $100$. As seen in Fig.~\ref{fig:coherent_mech}(a-b), distinguishing the ground and excited states requires a sufficiently large separation between the displaced resonator states. Achieving such a separation typically requires a longer evolution time, which in turn increases the effects of decoherence. Therefore, a desirable implementation would employ a large longitudinal coupling strength, enabling the required displacement to be generated within a shorter interaction time.

\subsubsection{Generation of cat states in a mechanical resonator}
The longitudinal interaction can also be utilized for generating cat states of the mechanical resonator.
A cat state is a quantum superposition of two coherent states with opposite phases.
For generation of cat states, we again consider unitary evolution under the Hamiltonian Eq. \eqref{H^I_Tm} 
with the initial state
$|\psi_0\rangle =  (|g\rangle + |e\rangle)  |0\rangle/\sqrt{2}.$

Since the Hamiltonian is time-independent, the time-evolution operator is
\begin{equation}
\hat{U}(t) = \exp\left(-\frac{i}{\hbar} \hat{H} t\right) = \exp\left(-i G_0 t (\hat{b} + \hat{b}^\dagger) \hat{\sigma}_z \right).
\end{equation}

This operator can be identified as a displacement operator conditioned on the qubit state:
\begin{equation}
\label{eq:beta}
\hat{U}(t) = \hat{D}(\beta \hat\sigma_z), \quad \text{with} \quad \beta = -i G_0 t,
\end{equation}
where \( \hat{D}(\beta) = \exp(\beta \hat{b}^\dagger - \beta^* \hat{b}) \) is the bosonic displacement operator.

Applying \( U(t) \) to the initial state:
\begin{align}
|\psi(t)\rangle &= \hat{U}(t) |\psi_0\rangle \\
&= \frac{1}{\sqrt{2}} \left( \hat{D}(\beta \hat\sigma_z)|g\rangle  |0\rangle + \hat{D}(\beta \hat\sigma_z)|e\rangle  |0\rangle \right) \\
&= \frac{1}{\sqrt{2}} \left( |g\rangle  \hat{D}(-\beta)|0\rangle + |e\rangle  \hat{D}(\beta)|0\rangle \right).
\end{align}

Since \( \hat{D}(\beta)|0\rangle = |\beta\rangle \), the coherent state with amplitude \( \beta \), we get
\begin{equation}
\label{eq:bell_cat}
|\psi(t)\rangle = \frac{1}{\sqrt{2}} \left( |g\rangle  |-\beta\rangle + |e\rangle  |\beta\rangle \right),
\end{equation}
where \( \beta = -i G_0 t \).

This is an entangled state between the qubit and the mechanical resonator, where each qubit state is associated with a coherent state of the resonator with opposite amplitudes.
By applying another \( \pi/2 \) rotation to the qubit, the state becomes:
\begin{equation}
    \label{eqn:Cat_state}
    |\Psi\rangle = \frac{1}{2} \left[ (|-\beta\rangle + |\beta\rangle)\, |g\rangle + (|\beta\rangle - |-\beta\rangle)\, |e\rangle \right].
\end{equation}
Here, the state is properly normalized, $\langle \Psi | \Psi \rangle = 1$. A projective measurement of the qubit in the excited (ground) state projects the resonator onto the odd (even) cat state: \(
|\psi_{\text{cat}}^{\pm}\rangle \propto |\beta\rangle \pm |-\beta\rangle.
\) with probability $P_e=(1-e^{-2|\beta|^2})/2$ ($P_g=(1+e^{-2|\beta|^2})/2$). In the limit $|\beta|\gg 1$, one obtains $P_e=P_g=1/2$.
This is shown in Fig. \ref{fig:coherent_mech}(c-d).

The cat state generation suffers from decoherence, whose rate is proportional to $\exp(-2|\beta|^2G_0t)$~\cite{Myatt2000lv}. As can be seen from Eq.~\eqref{eq:beta}, $\beta$ is dependent on time. Therefore, as time increases, the magnitude of $\beta$ increases while the coherence degrades. Thus, there is a trade-off: while larger cat states are favourable because they make the quantum superposition more distinct, at the same time, larger cat states lead to accelerated decoherence. Cat states of size $\approx 2$ have been demonstrated with fidelities of $\approx 70$-$80\%$~\cite{16microgram,twoboxes,Myatt2000lv}.
Furthermore, the cat-state generation time must be shorter than the qubit decoherence time, as decoherence destroys the phase coherence of the state given by Eq.~\eqref{eq:bell_cat}. The qubit measurement fidelity also directly affects the visibility of the projected cat state. 

The longitudinal coupling generally arises from flux-induced interactions with the qubit, and the mechanical resonators involved are typically in the MHz frequency range. These resonators require additional ground-state cooling since the protocol requires the mechanical resonator to be initially prepared in its vacuum state. 
Therefore, compared to cat-state generation schemes involving GHz-frequency resonators, MHz resonators require additional cooling and are more susceptible to degradation from thermal noise during the protocol.

Cat-state generation involving GHz-frequency mechanical resonators can be realized using acoustic modes coupled to a superconducting qubit through its charge degree of freedom. In this case, the interaction is mediated by transverse coupling. By operating in the dispersive regime, where the qubit is sufficiently detuned from the mechanical resonator, the effective qubit-resonator interaction takes the form $\propto \hat{b}^{\dagger}\hat{b}\,\hat{\sigma}_z$. By initially preparing the qubit in a superposition state and the mechanical resonator in a displaced coherent state, a Bell-cat state similar to that given in Eq.~\eqref{eq:bell_cat} can be generated. Subsequently, by following the same protocol discussed above, even and odd mechanical cat states can be generated.

\subsubsection{Quantum sensing with cat states}

Cat states constitute an important resource for quantum-enhanced metrology \cite{RevModPhys.90.035005}. The interference fringes of cat states are sensitive to external perturbations, allowing physical effects to be converted into measurable changes in the quantum interference pattern. Below, we discuss two particular sensing applications of cat states.

\textbf{Weak force detection with cat states:}

Consider the cat state $\vert \psi^{+}_{\mathrm{cat}}\rangle$ generated from the protocol discussed above. Let a weak classical force $\eta(t)$ (real) act on a mechanical resonator. This results in the displacement of the cat state into \cite{PhysRevA.66.023819}
\begin{equation}
D(\eta)|\psi^{+}_{\mathrm{cat}}(t)\rangle
=\frac{1}{\sqrt{2}} \left(
e^{i\theta}|\beta\rangle
+e^{-i\theta}|-\beta\rangle\right),
\label{displaced cat state sensing}
\end{equation}
where $\theta=2\,\mathrm{Im}(\beta \eta^{*})$ is the rotation parameter that captures the force-induced displacement, and
$
D(\eta) = e^{\eta(b^{\dagger}-b)}$ is the displacement operator.
If we substitute $|\pm\rangle=\frac{1}{\sqrt{2}}(|\beta\rangle\pm|-\beta\rangle)$ into Eq.~\eqref{displaced cat state sensing}, the resulting state can be interpreted as a qubit-like rotation between the even and odd cat states induced by the weak force. Here, $\vert\pm\rangle$ act as two orthogonal states analogous to the ground and excited states of a qubit.
For the case where $\beta$ is purely imaginary, $\theta=-|\beta|\eta$ is maximized. This relation shows that a small change in the rotation angle $\theta$ with respect to $\eta$ is proportional to the amplitude, or equivalently, the size of the cat state.
Equivalently, as shown in \cite{PhysRevA.66.023819}, the force sensitivity $\delta \epsilon$ scales as
\begin{equation}
    \delta \epsilon \ge \frac{1}{2\sqrt{\bar{n}}}
    \label{force sensitivity scale}
\end{equation}
where $\bar{n}=|\beta|^2$ is the mean boson number with $|\beta|^2=\bar{n}$, and this is the standard quantum limit (SQL) scaling for a single cat state.
This sensitivity can further be increased by incorporating multipartite entangled cat state and performing joint measurement to achieve Heisenberg-limited sensing \cite{PhysRevA.66.023819,RevModPhys.90.035005}.
For an $N$-mode entangled cat state with total photon number $n_{\mathrm{tot}}=N|\beta|^{2}$, the uncertainty in this force parameter is bounded by $\delta \epsilon \propto 1/\sqrt{Nn_{\mathrm{tot}}}$.
This offers a $\sqrt{N}$ sensitivity enhancement over independent single-mode cat states. 
The generation of multipartite cat states in hybrid electromechanical systems has been proposed in Ref.~\cite{PhysRevA.110.023726}. The scheme employs two qubit-mechanical subsystems, where a Bell-state measurement performed on the superconducting qubits realizes entanglement swapping, thereby projecting the two mechanical resonators onto a bipartite entangled cat state. In addition, the generation of multipartite cavity cat states in qubit-cavity systems has also been demonstrated in Ref.~\cite{flyingcat}.

Since force and acceleration are related, the displacement-based sensing framework can be directly adapted for acceleration sensing with comparable estimation precision. Nevertheless, the practical realization of such schemes is severely constrained by the extreme susceptibility of cat states to decoherence.

\textbf{Frequency shift sensing with cat states:}
Beyond displacement sensing, Schrödinger cat states are considered as a resource for quantum-enhanced interferometry in single bosonic modes, enabling phase estimation with sensitivities approaching the Heisenberg limit.
A frequency shift induces a phase shift in the quantum state, corresponding to a rotation in phase space. As discussed in \cite{ng3t-p1n1}, in the rotating frame, the evolution of the system under a frequency shift $\epsilon$ relative to the reference frequency is governed by the Hamiltonian
\(
H_{\mathrm{FS}}=\hbar\epsilon\, b^{\dagger}b,
\) where $\hbar$ is the reduced Planck constant. After an evolution time $T$, the accumulated phase is $\theta=-\epsilon T$. Consequently, an initial cat state $\mathcal{N}(\ket{0}+\ket{\alpha})$ evolves into $\mathcal{N}(\ket{0}+\ket{e^{i\theta}\alpha})$, where $\mathcal{N}$ is the normalization constant. The accumulated relative phase,
$\phi=\operatorname{arg}\!\left(\langle\alpha|e^{i\theta}\alpha\rangle\right)
=|\alpha|^2\sin\theta,
$ scales linearly with the mean boson number $\bar{n}=|\alpha|^2$. For small frequency shifts, $\phi\simeq\bar{n}\theta$, enabling Heisenberg-limited frequency estimation under optimal measurement conditions ~\cite{ng3t-p1n1}. Since the resonance frequency of a mechanical oscillator depends on its effective mass, the adsorption of a small mass produces a measurable frequency shift. Consequently, the same cat-state interferometric protocol developed for frequency sensing can be directly extended to mass sensing, with the accumulated phase providing an amplified signature of the added mass.

As discussed above, despite their significant metrological potential, the practical realization of cat-state-based quantum sensing remains experimentally challenging. The quantum advantage offered by cat states originates from the coherence between their macroscopically distinct components. However, decoherence rapidly degrades the interference fringes that give rise to the enhanced sensitivity, and this effect becomes increasingly severe as cat states grow larger. In mechanical resonators, thermal noise poses an additional challenge by destroying the phase coherence of mechanical cat states. Nevertheless, ongoing advances in high-coherence quantum platforms, cryogenic operation, quantum control techniques, and bosonic error-correction and stabilization protocols continue to improve the robustness of cat states, bringing practical quantum-enhanced sensing closer to experimental realization.

\begin{figure}[!hbt]
    \centering
    \includegraphics[width=90mm]{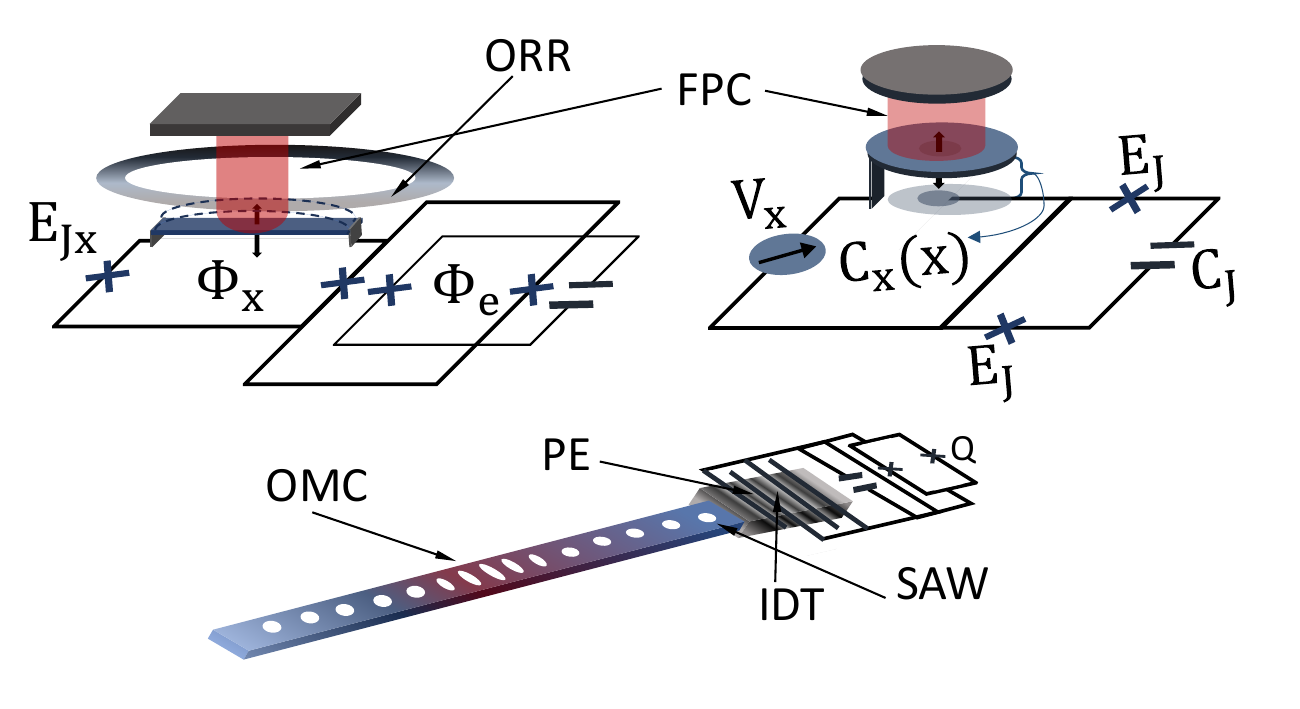}
    \caption{Schematic of the Electro-Optomechanical System: For a suspended mechanical beam (top left), the optical mode can be coupled to the mechanical mode by either forming a Fabry-P\'{e}rot
    cavity (FPC) with the beam acting as an end mirror or by placing an optical ring resonator (ORR) near the beam, where the optical resonator field picks up the vibrational signal from the beam. A suspended membrane also couples with the optical mode by forming a Fabry-P\'{e}rot
    cavity (top right). In an optomechanical crystal (OMC), the mechanical mode within the crystal couples with the confined optical mode through the photoelastic effect (bottom) (Adapted from \cite{Mirhosseini2020}. In all three cases, the qubit couples with the optical mode through mechanical-optical coupling. }
    \label{fig:system_qubit_mech_phase1}
\end{figure}

\subsubsection{Mechanical resonator-based force sensing}

Consider an unknown weak force acting on a mechanical resonator that is longitudinally coupled to a qubit, as described by Eq.~\eqref{Eq:Long}. The effect of this force can be incorporated into the Hamiltonian by adding a term of the form $\eta(b+b^\dagger)$, where $\eta$ characterizes the strength of the force.
Under the action of this force, the state of the qubit-mechanical system evolves in the interaction picture as \cite{suri2025}
\begin{equation}
\ket{\Psi(t)}
=
e^{-i\theta/2}
\frac{1}{\sqrt{2}}
\left(
e^{i\theta}\ket{+}\ket{\alpha_+(t)}
+
\ket{-}\ket{\alpha_-(t)}
\right),
\end{equation}
where $\ket{\pm}$ denote the eigenstates of the qubit operator $\sigma_x$, and $\ket{\alpha_\pm(t)}$ are coherent states of the mechanical resonator with amplitudes
$\alpha_\pm(t)=A_\pm\left(1-e^{i\omega_b t}\right),\qquad A_\pm=(\eta\pm G)/\omega_b,$
with $G$ denoting the qubit-mechanical coupling strength. The phase
$\theta(t)= 4\eta G\left(\omega_b t-\sin(\omega_b t)\right)/\omega_b^2$
represents the relative phase accumulated between the two qubit components.
In the oscillator phase space, the coherent states $\alpha_\pm(t)$ trace circular trajectories with radii determined by $A_\pm$. By tracing out the mechanical degree of freedom, the reduced density matrix of the qubit can be obtained. The expectation value of $\sigma_x$ is then given by $\langle \sigma_x \rangle=e^{-d(t)}\cos\!\bigl[\theta(t)\bigr],$ where $d(t)=8G^2\sin^2\!\left(\omega_b t/2\right)/\omega_b^2.$
The force signal $\eta$ is encoded solely in the phase $\theta(t)$, while the factor $e^{-d(t)}$ originates from the overlap of the two oscillator trajectories.
At time $t=2\pi/\omega_b$, the qubit and mechanical resonator become completely disentangled. Consequently, $d(t)=0$, and the measurement signal reduces to $\cos\!\bigl[\theta(t)\bigr]$.
At this point, the information about the force is entirely transferred to the qubit phase, allowing the signal to be retrieved with maximal visibility.

As proposed in \cite{suri2025}, the sensitivity to the force signal $\eta$ can be further enhanced by a protocol involving squeezing of the mechanical resonator together with the application of $\pi$-pulses on the qubit. After completion of the protocol, the state of the qubit-mechanical system at time $T=3\times 2\pi/\omega_b$ evolves to $\ket{\Psi(T)}=e^{i\theta_T/2}\hat{I}\ket{\Psi(0)},$
where $\ket{\Psi(0)}$ is the initial state and $\hat{I}$ is the identity operator. The accumulated phase is given by
$\theta_T=8\pi \eta G/\omega_b^2+4a_\pm,$
where $a_\pm=\pm8(e^r-1)G\eta/\omega_b^2$
is the area enclosed by the oscillator trajectory in phase space during the protocol. The second term therefore represents a geometric phase contribution proportional to the enclosed area. Since the area scales with the squeezing parameter $r$, the force signal is encoded in a geometric phase that is amplified by squeezing.
The enhancement can be roughly understood by examining the sensitivity of the measurement signal to the force parameter,
$\partial \langle \sigma_x \rangle/\partial \eta.$
For the squeezing-assisted protocol, the accumulated phase scales as $\theta_T \propto
(e^r-1)\eta,$ and therefore, for large squeezing $r$,$
\partial \theta_T/\partial \eta
\propto e^r.$ Consequently, the variation of the measured signal $\langle \sigma_x \rangle$ with respect to the force strength $\eta$ is enhanced by a factor of approximately $e^r$, leading to a substantial improvement in force-sensing sensitivity.
A more concrete quantity for measuring the signal is through the Fisher information as discussed in \cite{suri2025}.

\subsubsection{Comparison of Transmon and Fluxonium Electromechanical Systems}

An important feature that distinguishes transmon- and fluxonium-based electromechanical systems is the anharmonicity of their energy levels. 
In the transmon case, the anharmonicity is approximately uniform ($\alpha \approx -E_C$), whereas in the fluxonium case it is highly nonuniform and significantly larger, as shown in Fig.~\ref{fig:trans_spec}. 
The matrix elements responsible for transitions between energy levels are relatively simple in the transmon case, while in the fluxonium case they exhibit a much more complex structure.
In the transmon case, transitions are significant primarily between adjacent energy levels, while transitions between nonadjacent levels are strongly suppressed. In contrast, the fluxonium allows transitions between multiple energy levels, provided that the two states involved have opposite parity \cite{PhysRevX.11.011010}. For example, transitions between $|g\rangle$ and $|e\rangle$, and between $|g\rangle$ and $|h\rangle$, are allowed, whereas the transition between $|g\rangle$ and $|f\rangle$ is forbidden because the two states have the same parity.
Consequently, different transitions of the fluxonium can be engineered to couple to different physical systems more conveniently than in a transmon. For example, transitions such as $|g\rangle \leftrightarrow |h\rangle$ can be directly addressed, enabling one transition to be coupled to a mechanical resonator while another is coupled to a microwave resonator for qubit control and readout.

Another important consequence of the large anharmonicity of the fluxonium is that it can operate at transition frequencies in the few-MHz range \cite{PhysRevX.11.011010}. This is particularly advantageous for coupling to mechanical resonators, which typically have resonance frequencies in the MHz regime. Since the qubit and mechanical frequencies can be brought into resonance, it becomes possible to realize a strong transverse qubit-mechanical interaction without the need for additional parametric drives. 
Another advantage of the large anharmonicity of the fluxonium is that higher-energy transitions in the GHz range can be utilized for dispersive readout using a microwave cavity. This allows one to retain standard circuit-QED readout techniques even when the computational transition lies in the MHz regime.
A caveat is that a fluxonium qubit operating at MHz frequencies requires efficient ground-state initialization in order to observe quantum signatures, since the thermal population can become significant at dilution refrigerator temperatures. Nevertheless, such ground-state cooling and control of low-frequency fluxonium qubits have been demonstrated experimentally. Furthermore, schemes for the simultaneous cooling of both the fluxonium and the mechanical resonator, leveraging the rich multilevel structure of the fluxonium spectrum, have also been explored~\cite{Nongthombam2025}.

A comparatively larger body of work on electromechanical platforms has been explored using transmon qubits. This is primarily because the transmon architecture is a more mature and well-established superconducting qubit platform than fluxonium. In addition, the fabrication and experimental implementation of transmon devices are generally more straightforward than those of fluxonium circuits, which require the realization of large superinductances and more complex circuit designs.
Transmons also benefit from a well-developed circuit-QED ecosystem, including high-fidelity microwave control, dispersive readout, tunable couplers, and scalable multi-qubit architectures.

In both transmon- and fluxonium-based electromechanical systems, relaxation and decoherence arise from noise channels associated with both the charge and phase degrees of freedom \cite{PhysRevX.11.011010,transmon1}. Noise coupled through the charge degree of freedom includes coupling to transmission-line resonators, radiative relaxation through charge-control lines, and Purcell decay. Flux noise, on the other hand, couples through the phase degree of freedom and can lead to qubit dephasing, particularly in flux-sensitive operating regimes.
In addition, quasiparticles, generated through the breaking of Cooper pairs, can induce both relaxation and dephasing in Josephson-junction-based qubits. Another important source of decoherence is dielectric loss originating from insulating materials associated with circuit capacitors. 
The presence of noise can induce fluctuations in both the phase and charge degrees of freedom of the qubit, as well as in the mechanical resonator itself. As an example, consider a fluctuating flux noise $\delta\Phi(t)$ acting on the fluxonium. The inductive term of the fluxonium-mechanical Hamiltonian becomes
\begin{equation}
    \hat{H}_L=
    \frac{E_L}{2}
    \left(
    \hat{\theta}^{\mathrm F}
    -
    \frac{2\pi(\Phi_e+\hat{\Phi}_b+\delta\Phi(t))}
         {\Phi_0}
    \right)^2 ,
\end{equation}
where $\Phi_e$ is the external bias flux and $\hat{\Phi}_b$ denotes the flux generated by the mechanical resonator. Expanding the Hamiltonian yields
\begin{align}
\hat H_L
=&\,
\frac{E_L}{2}
\left(
\hat{\theta}^{\mathrm F}
-\frac{2\pi(\Phi_e+\delta\Phi(t))}{\Phi_0}
\right)^2
+
E_L
\hat{\theta}^{\mathrm F}\frac{2\pi{\hat\Phi_b}}{\Phi_0}
-E_L\hat\Phi_b\delta\Phi(t)
\left(\frac{2\pi}{\Phi_0}\right)^2+\cdots
\end{align}

The second term describes the coherent fluxonium-mechanical interaction. The first term corresponds to the direct coupling of flux noise to the phase degree of freedom which leads to dephasing. The third term represents a noise-induced modulation of the mechanical interaction, coupling the mechanical resonator to the fluctuating flux environment.
The noise source may be treated either as a classical stochastic process, $\delta\Phi(t)$, or as a quantum operator, $\delta\hat{\Phi}(t)$, depending on the physical origin of the environment. In an open-system description, the effect of these noise sources is captured by the dissipation and dephasing rates entering the Lindblad master equation. Consequently, the observation of coherent quantum signatures of the fluxonium-mechanical interaction requires the coherent coupling strength to exceed the relevant decoherence and dissipation rates.
Analogous to the flux-coupled case, in a charge-coupled electromechanical system the fluctuations induced in the mechanical resonator arise from charge noise.

\section{Optomechanical Coupling with Qubit}
\label{sec:Optomechanical Coupling with Qubit}

In this section, we discuss the integration of an optical cavity into the qubit–mechanical system, forming a hybrid electro-optomechanical system (see Fig.~\ref{fig:system_qubit_mech_phase1}). 

Depending on the type of qubit–mechanical system described in the previous section, the optical component can be integrated in different ways. 
In both coupling schemes—i.e., coupling via the charge degree of freedom or via the flux degree of freedom—the displacement of the mechanical resonator alters the resonant frequency of the optical cavity \( \omega_a \), resulting in a mechanical–optical interaction term. Expanding the frequency shift of the optical cavity up to linear order in displacement, we obtain \( \omega_c(x) = \omega_a + (\partial\omega_c/\partial x) x \)    \cite{RevModPhys.86.1391}. Expressing the displacement in terms of the creation and annihilation operators \( \hat{b}^\dagger \) and \( \hat{b} \), the Hamiltonian of the full electro-optomechanical system (for both transmon and fluxonium implementations) reads \cite{PhysRevA.104.023509},
\begin{align}
    \label{eqn:electro-optomech}
    \hat{H}_{elo} &= \hbar\frac{\Omega_{Tx}}{2} \hat\sigma_z + \hbar\omega_b \hat{b}^\dagger \hat{b} + \hbar G_L (\hat{b}^\dagger + \hat{b})\hat\sigma_z \notag \\&\quad
    + \hbar\omega_a \hat{a}^\dagger \hat{a} + \hbar g \hat{a}^\dagger \hat{a} (\hat{b}^\dagger + \hat{b}),
\end{align}
where \( g = X_{\text{zpf}}\, \partial\omega_c/\partial x \) is the single-photon optomechanical coupling strength. 
Here, longitudinal coupling between the qubit and the resonator is considered. The shift in the cavity frequency arises from the back-action of the radiation pressure force exerted by the cavity photons on the mechanical resonator~\cite{braginski1967,braginsky1968,braginsky1995quantum, PhysRevLett.51.1550}. This force is given by
\(
    \hat{F} = -\frac{\partial \hat{H}_{elo}}{\partial x} = \left(\hbar g/X_{\text{zpf}}\right) \hat{a}^\dagger \hat{a}.
\)
The force depends on the photon number in the cavity; therefore, the more photons present, the greater the force imparted on the resonator, resulting in a stronger coupling. 
In typical experiments, the single-photon coupling strength \( g \) is on the order of 1~Hz (suspended mechanical resonator)—much smaller than the other frequency scales in the system. 
However, since the radiation pressure force (and hence the coupling strength) increases with the photon number, the optomechanical coupling can be enhanced by coherently driving the cavity. 
To analyse this effect, we first write down the Langevin equations for both the mechanical resonator and the cavity.

\begin{subequations}
    \begin{eqnarray}
    \label{eqn:langevin_dadt}
    \hspace{-0.6cm} \dot{\hat{a}} &=& -\left(\frac{\kappa + \kappa_1}{2}+i\omega_a\right)\hat{a}-ig~\hat{a}(\hat{b}^\dagger+\hat{b}) \notag \\ && + \sqrt{\kappa_1}\hat{a}_{in1} +\sqrt{\kappa_2}\hat{a}_{in2},\\
    \hspace{-0.6cm} \dot{\hat{b}} &=& -\left(\frac{\gamma}{2}+i\omega_b\right)\hat{b}-iG_L\hat\sigma_z -ig\hat{a}^\dagger \hat{a} + \sqrt{\gamma}\hat{b}_{in}.
    \end{eqnarray}     
\end{subequations}
Here, $\kappa$, $\kappa_1$ and $\kappa_2$ are the decay rates corresponding to the intra cavity field, input field $a_{in1}$ and intrinsic noise field $\hat{a}_{in2}$. $\hat{b}_{in}$ is the input noise operator of the resonator. 
The cavity is driven through the input port $a_{in1}$. When the drive is strong, the input field can be separated into the quantum and classical parts: $\hat{a}_{in1}\rightarrow \hat{a}_{in1}+A(t)$, where $A(t)$ is the amplitude of the drive. Substituting in Eq. \eqref{eqn:langevin_dadt}, we have
\begin{align}
    \label{eqn:dadt}
    \dot{\hat{a}} = -\left(\frac{\kappa + \kappa_1}{2}+i\omega_a\right)\hat{a}-ig~\hat{a}(\hat{b}^\dagger+\hat{b}) \notag \\ + \sqrt{\kappa_1}(A(t)+\hat{a}_{in1}) +\sqrt{\kappa_2}\hat{a}_{in2}.
\end{align}

This change can be implemented in the Hamiltonian Eq. \eqref{eqn:electro-optomech} by adding a drive term $\hat{H}_d = \hbar \sqrt{\kappa_1}A(t)(\hat{a}\,exp(-i\omega_d t) + \hat{a}^\dagger exp(i\omega_d t))$. 
The optical frequency is typically large (in THz) compared to the qubit and mechanical resonator and the drive term is time dependent. So, it is advisable to go into the drive frame. 
Therefore, we have
\begin{eqnarray}
    \label{eqn:electro-optomech1}
    \hat{H}_{elo} &=& \hbar\frac{\Omega_{Tx}}{2}\hat\sigma_z+\hbar\omega_b \hat{b}^\dagger \hat{b}+\hbar G_{L}(\hat{b}^\dagger+\hat{b})\hat\sigma_z  \nonumber \\ &&+ \hbar \hat{a}^\dagger \hat{a}\Delta_a + \hbar g \hat{a}^\dagger \hat{a}(\hat{b}^\dagger + \hat{b}) + \hbar\epsilon(t)(\hat{a}^\dagger + \hat{a}),
\end{eqnarray}
where $\Delta_a = \omega_a-\omega_d$ and $\epsilon(t)=\sqrt{\kappa} A(t)$.
Since the drive is strong, the cavity is coherently populated with a large number of photons, and the system dynamics lean toward the classical regime. 
To isolate the quantum part, we separate the cavity field into classical $(\alpha)$ and quantum components: $\hat{a}\rightarrow \hat{a}+\alpha$. Similarly, the mechanical motion can be separated into quantum and classical $(\beta)$ components:$\hat{b}\rightarrow \hat{b}+\beta$. 
The classical part of the mechanical motion is induced by the radiation pressure force exerted by the cavity photons. 
This separation into classical and quantum parts can be understood as the classical part representing a coherent displacement in phase space, while the quantum part represents small fluctuations around the displaced amplitude. 
The coherent amplitudes depend on the photon (phonon) number as $\alpha=\sqrt{n_a}$ $(\beta=\sqrt{n_m})$, $n_a$ ($n_m$) is the average photon (phonon) number.  
The quantum part of the Langevin equation, obtained after the classical-quantum separation, is given by
\begin{subequations}
    \begin{eqnarray}
    \label{eqn:quantum_langevin_dadt}
    \dot{\hat{a}} &=& -\left(\frac{\kappa + \kappa_1}{2}+i\Delta_a+ig(\beta+\beta^*)\right)\hat{a}-ig~\hat{a}(\hat{b}^\dagger+\hat{b})  \nonumber \\  && -ig\alpha(\hat{b} + \hat{b}^\dagger) + \sqrt{\kappa_1}\,\hat{a}_{in1} +\sqrt{\kappa_2}\,\hat{a}_{in2},\\
    \dot{\hat{b}} &=& -\left(\frac{\gamma}{2}+i\omega_b\right)\hat{b}-iG_L\hat\sigma_z -ig(\alpha \hat{a}^\dagger+\alpha^*\hat{a}) \notag \\ &&-ig\hat{a}^\dagger \hat{a} + \sqrt{\gamma}\,\hat{b}_{in}
    \end{eqnarray}     
\end{subequations}
and the classical part by
\begin{subequations}
 \label{eqn:classical_langevin_dadt}
    \begin{eqnarray}
    \hspace{-0.4cm}
    \dot{\alpha} &=& -\left(\frac{\kappa + \kappa_1}{2}+i\Delta_a+ig(\beta+\beta^*)\right)\alpha -i\epsilon,\\
    \hspace{-0.4cm}\dot{\beta} &=& -\left(\frac{\gamma}{2}+i\omega_b\right)\beta -ig\alpha\alpha^* 
    \end{eqnarray}     
\end{subequations}
In Eq. \eqref{eqn:quantum_langevin_dadt}, the linear term $-ig\alpha(\hat{b} + \hat{b}^\dagger)$ $(-ig(\alpha \hat{a}^\dagger+\alpha^*\hat{a})\,)$ is amplified by the amplitude $\alpha$ compared to the nonlinear term $-ig~\hat{a}(\hat{b}^\dagger+\hat{b})$ $(-ig~\hat{a}^\dagger \hat{a})$. 
Therefore, we can ignore the nonlinear term and retain only the linear term. This is known as linearization. The corresponding linearized Hamiltonian can be written as 
\begin{eqnarray}
    \label{eqn:electro-optomech_linear}
    \hat{H}_{elo} &=& \hbar\frac{\Omega_{Tx}}{2}\hat\sigma_z+\hbar\omega_b \hat{b}^\dagger \hat{b}+\hbar G_{L}(\hat{b}^\dagger+\hat{b})\hat\sigma_z  \nonumber \notag \\ && +\hbar \hat{a}^\dagger \hat{a}\Delta_a + \hbar G_\alpha(\hat{a}^\dagger+\hat{a})(\hat{b}^\dagger + \hat{b}),
\end{eqnarray}
where we have merged $\Delta_a\rightarrow\Delta_a+g(\beta+\beta^*)$ for simplicity. 
Upon substituting the classical and quantum parts, an extra term, $\hbar G_{L}(\beta^*+\beta)\hat{\sigma_z}$ arises from the third term of Eq. \eqref{eqn:electro-optomech1}. 
Although this extra term is not visible in the Langevin equation Eq. \eqref{eqn:quantum_langevin_dadt}, it shows up in the qubit's dynamics. Since this term only shifts the qubit frequency, it can be absorbed into, $\Omega_{Tx}\rightarrow\Omega_{Tx}+G_{L}(\beta^*+\beta)$. 
A similar linearized Hamiltonian Eq.\eqref{eqn:electro-optomech_linear} can be derived for a transverse electro-optomechanical system by replacing the third term with the transverse interaction term:
\begin{eqnarray}
    \label{eqn:electro-optomech_linear_trans}
    \hat{H}'_{elo} &=& \hbar\frac{\Omega_{Tx}}{2} \hat\sigma_z + \hbar\omega_b \hat{b}^\dagger \hat{b}  + \hbar \hat{a}^\dagger \hat{a} \Delta_a + \hbar G_{tm}(\hat{b}^\dagger + \hat{b})\hat\sigma_x  \notag\\
     &&  + \hbar G_{tm}(\beta^* + \beta)\hat\sigma_x + \hbar G_\alpha (\hat{a}^\dagger + \hat{a})(\hat{b}^\dagger + \hat{b}).
\end{eqnarray}
A weak coherent drive of the qubit is induced from the mechanical displacement term \( \hbar G_{tm}(\beta^* + \beta)\sigma_x \), which can be neglected.
Under the rotating wave approximation, we get a Jaynes–Cummings and beam-splitter interaction Hamiltonian:
\begin{eqnarray}
    \label{eqn:electro-optomech_linear_trans_RWA}
    \hat{H}'_{elo} &=& \hbar\frac{\Omega_{Tx}}{2} \hat\sigma_z + \hbar\omega_b \hat{b}^\dagger \hat{b} + \hbar \hat{a}^\dagger \hat{a} \Delta_a + \hbar G_{tm}(\hat{b}^\dagger \hat\sigma_- + \hat{b} \hat\sigma_+)\nonumber \\
     && +  \hbar G_{tm}(\beta^* + \beta)\hat\sigma_x + \hbar G_\alpha (\hat{a}^\dagger \hat{b} + \hat{a} \hat{b}^\dagger).
\end{eqnarray}

Below, we discuss some applications of both transverse and longitudinal coupling.

\subsubsection{State transfer from a qubit to an optical photon (quantum transduction)}
\begin{figure}[!hbt]
    \centering
    \includegraphics[width=40mm]{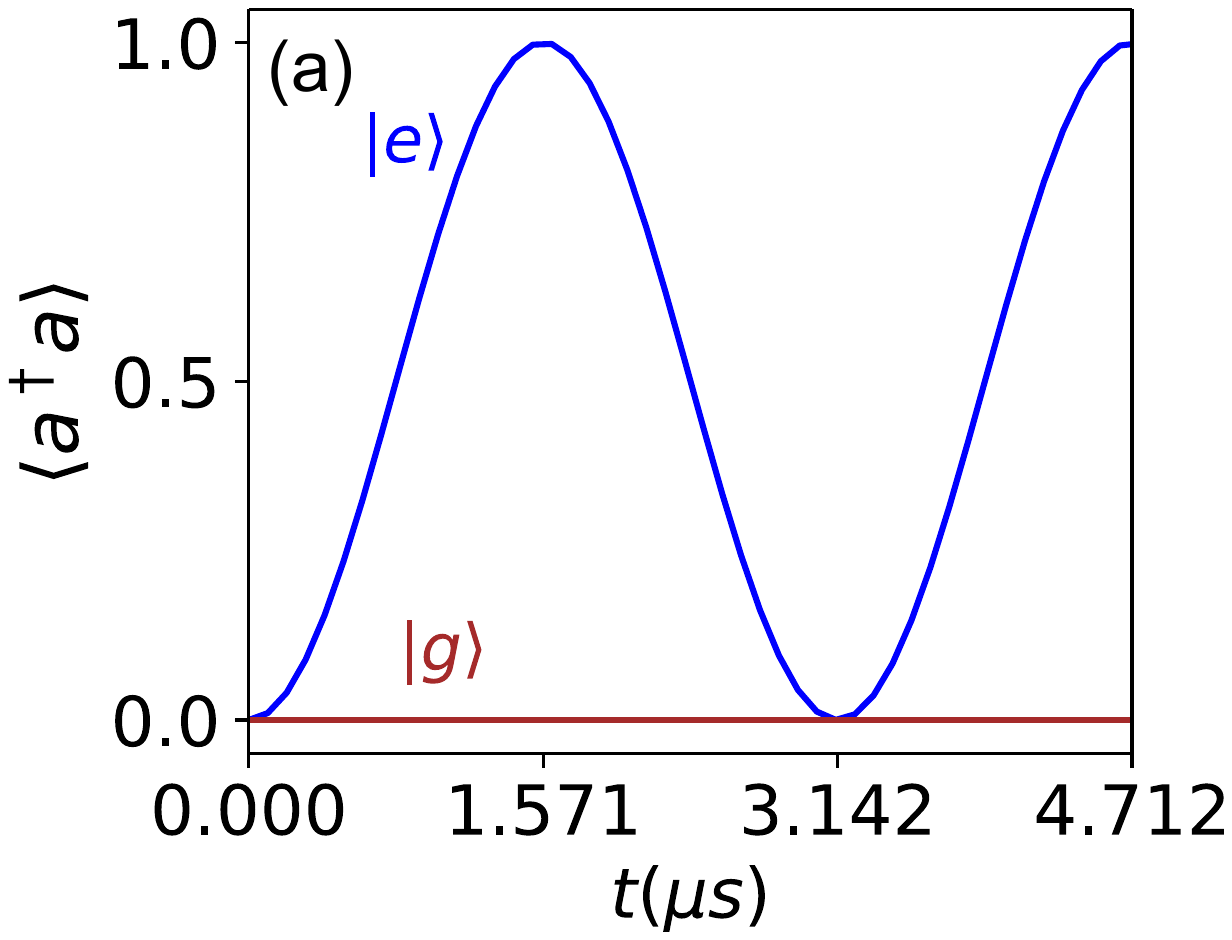}
    \vspace{-12mm}
    \includegraphics[width=40mm]{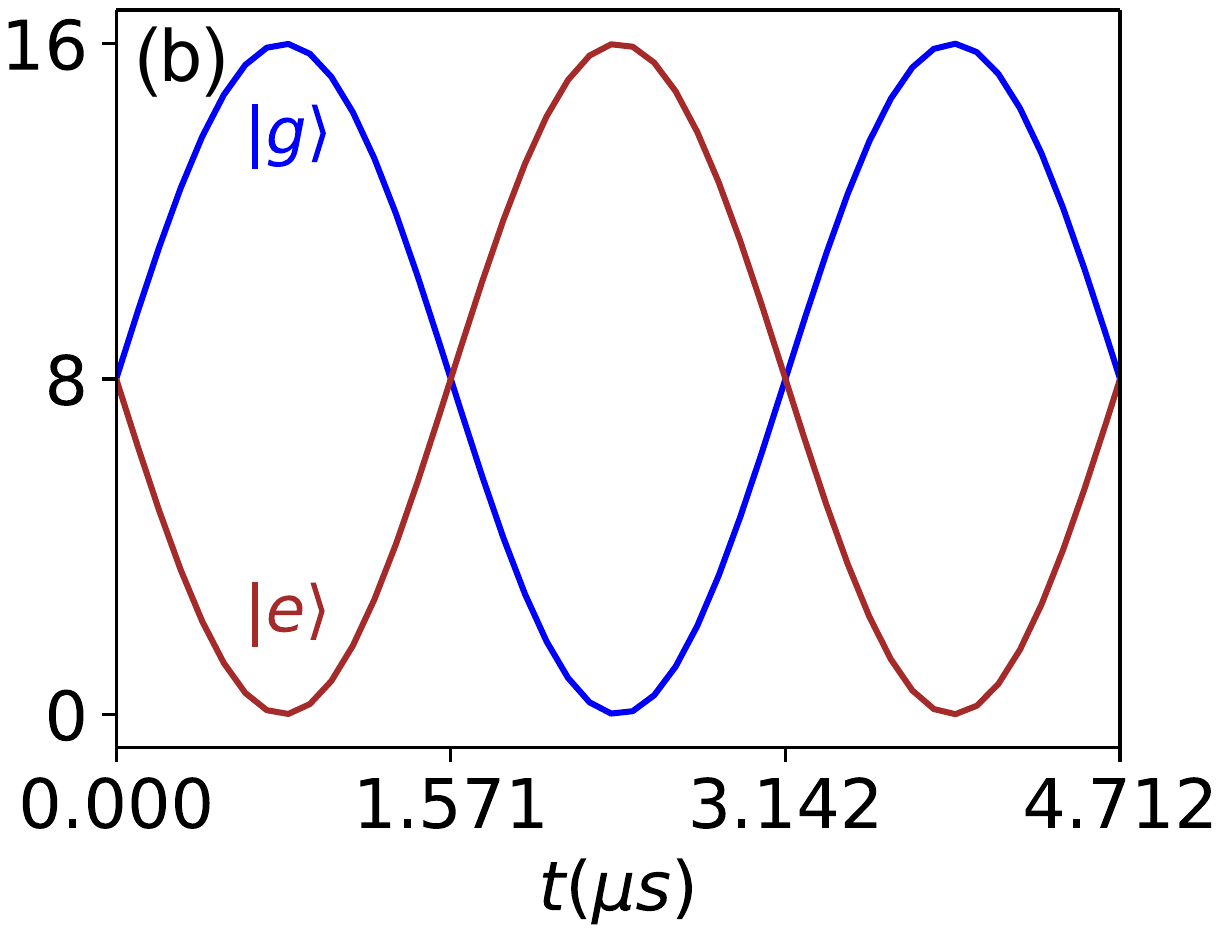}
    \vspace{-12mm}
    \caption{
    State transfer between the qubit and the optical photon:  
    (a) when the qubit is transversely coupled to the mechanical resonator, and  
    (b) when the mechanical resonator is longitudinally coupled to the qubit.  
    Parameters used:  
    (a) $G_\alpha = 1~\text{MHz}$, optical photon initially in the ground state;  
    (b) $G_\alpha = 1~\text{MHz}$, optical photon initially in a coherent state with amplitude equal to the absolute value of the coherently excited amplitude of the resonator. Adapted from \cite{PhysRevA.108.043501}.
    }
    \label{fig:optical_state_transfer}
\end{figure}

The transverse interaction shown in Eq.~\eqref{eqn:electro-optomech_linear_trans_RWA} is ideal for transferring the qubit state to the optical photon, as both the qubit–mechanical and mechanical–optical couplings facilitate state swapping. 
The full state transfer is realized through a double-swapping scheme, where the qubit state is first mapped onto the mechanical resonator and subsequently transferred to the optical mode.
To implement this, we initially focus on the qubit–mechanical coupling while neglecting the mechanical–optical interaction  \cite{Mirhosseini2020,PhysRevLett.108.153603, PhysRevA.108.043501}. 
The process of state swapping between the qubit and the resonator has been discussed earlier. 
Specifically, the initial state \( \lvert e,\, 0_m \rangle  \) evolves to \(\lvert g, 1_m\rangle \) under resonant Jaynes–Cummings interaction.
Once this swap is complete, the qubit–mechanical coupling can be effectively turned off by detuning the qubit from the mechanical resonator. Subsequently, the linear beam-splitter-type mechanical–optical coupling (last term in Eq. \eqref{eqn:electro-optomech_linear_trans_RWA}) is activated by driving the optical cavity. This interaction facilitates the transfer of excitation from the mechanical resonator to the optical cavity.
This second state-swap process is illustrated in Fig.~\ref{fig:optical_state_transfer}, where the mechanical–optical state evolves from \(\lvert 1_m, 0_p \rangle \) to \(\lvert 0_m, 1_p \rangle \) at interaction times \( t = \frac{(2n+1)\pi}{G_\alpha} \), with \( n = 0,1, 2, 3, \ldots \).

We have seen that the longitudinal coupling between the qubit and the resonator can be used to encode the qubit information into the coherent excitation state of the resonator. 
Using the hybrid electro-optomechanical Hamiltonian~\eqref{eqn:electro-optomech_linear}, this encoded information can be transferred from the mechanical mode to the optical mode.
We begin by focusing on the qubit–mechanical coupling while neglecting the mechanical–optical interaction, assuming that the optical cavity is not driven and that \( g \ll G_L \), or by detuning the qubit away from the cavity frequency. 
As discussed earlier, the qubit state is encoded in the mechanical resonator by evolving under the longitudinal qubit–mechanical interaction. 
This interaction is then switched off—for example, by tuning the external flux—and the optical cavity is driven to activate the beam-splitter interaction between the mechanical resonator and the optical mode. The qubit state can subsequently be inferred by measuring the optical cavity.
The evolution of the average photon number in the optical cavity is shown in Fig.~\ref{fig:optical_state_transfer}. 
The figure shows that measuring the average cavity photon number at time intervals \( t = \pi/2G_\alpha \), starting from \( t = \pi/4G_\alpha \), allows one to distinguish the qubit state based on the presence or absence of photons.
If photons are detected at times \( t = (2n+1)\pi/4G_\alpha \), where \( n = 0, 2, 4, \ldots \), it indicates that the qubit was initially in the ground state. Conversely, if no photons are detected at these same intervals, the qubit was in the excited state.
Similarly, if photons are detected at times \( t = (2n+1)\pi/4G_\alpha \), where \( n = 1, 3, 5, \ldots \), this corresponds to the qubit being initially in the excited state, while their absence at these intervals suggests the qubit was in the ground state.

In quantum transduction protocols, such as the excitation-swap process that converts a qubit excitation into an optical excitation, several requirements must be met for successful experimental implementation. These include frequency matching, large transduction bandwidth, high conversion efficiency, and low added noise. While experiments have achieved microwave-to-optical conversion efficiencies as high as $\sim 40\%$, such performance has generally been accompanied by significant added noise of approximately $50$ quanta~\cite{andrews2014bidirectional,Higginbotham2018cn}. Conversely, near-quantum-limited operation with added noise below one quantum has been demonstrated, but only at substantially lower conversion efficiencies, typically on the order of $10^{-5}$~\cite{Mirhosseini2020}. There appears to be a trade-off between high conversion efficiency and low added noise, which remains a significant challenge for the experimental realization of quantum transducers. Systems exhibiting low added noise generally employ electro-optomechanical architectures based on GHz-frequency mechanical acoustic resonators fabricated in crystalline materials. The low noise performance is primarily attributed to the high mechanical frequency, which corresponds to an almost ground-state thermal occupation at dilution refrigerator temperatures ($T \sim 15~\mathrm{mK}$).
Furthermore, the single-photon optomechanical coupling strength in GHz resonators can reach values of several hundred of kHz and can be further enhanced through optical pumping \cite{Mirhosseini2020}. Since the intrinsic single-photon coupling is already large, relatively modest optical drive powers are sufficient to achieve strong photon-enhanced optomechanical coupling. Lower optical pump powers are advantageous because they reduce optical absorption-induced heating and consequently suppress added noise.
In contrast, MHz-frequency mechanical resonators typically exhibit single-photon optomechanical coupling strengths in the range of a few Hz \cite{andrews2014bidirectional}, which are several orders of magnitude smaller than those achievable in GHz nanomechanical resonators. As a result, significantly higher optical pump powers are required to enhance the effective optomechanical coupling. The increased optical drive leads to stronger optical absorption, heating, and excess noise.
Another important source of added noise in MHz resonators is the large thermal phonon occupation, which remains significant even at dilution refrigerator temperatures. In contrast, GHz mechanical resonators are nearly in their quantum ground state under the same conditions.
The weaker single-photon optomechanical coupling in MHz resonators primarily arises from their larger effective mass, larger mode volume, and weaker optical and mechanical confinement, which result in a smaller optical frequency shift per displacement and a large frequency mismatch compared to GHz nanomechanical resonators.
In addition, there are several other practical considerations in implementing qubit-to-optical quantum transduction systems. These include quasiparticle generation due to optical absorption, which is one of the challenges in integrating optical and superconducting quantum systems; phonon leakage, which lowers the overall conversion efficiency; parasitic modes, which introduce additional loss channels; material losses arising from dielectric defects and two-level-system (TLS) defects; compatibility between piezoelectric films and superconducting circuits; and decoherence of the qubit induced by the coupled mechanical and optical subsystems.

Although achieving both high conversion efficiency and low added noise remains a significant challenge, experimental demonstrations of transduction from a single qubit excitation to an optical photon have been realized in piezoelectrically coupled electro-optomechanical systems operating in the low-efficiency, low-added-noise regime \cite{Mirhosseini2020}. Another important trade-off discussed in this work is the increase in the thermal occupation of the mechanical resonator with increasing optical measurement pulse duration. While a longer measurement pulse improves the efficiency of optical signal detection, it also leads to increased heating and thermal population of the mechanical mode. Consequently, the measurement pulse duration must be optimized to balance efficient signal detection against the detrimental effects of thermal noise.

\subsubsection{Ground-state cooling of the mechanical resonator using an electro-optomechanical system}

The electro-optomechanical system can be utilized for hybrid ground state cooling of the mechanical resonator. The cooling of the resonator is achieved by applying a red-detuned drive to both the qubit and the optical cavity. Due to this red detuning, the qubit is excited to a higher energy level by absorbing incoming photons from the external drive as well as phonons from the mechanical oscillator. As a result, the mechanical resonator loses phonons to the qubit, leading to cooling. Simultaneously, optical photons enter the cavity by absorbing phonons from the oscillator, further contributing to the cooling process. These two processes occur concurrently, enabling ground state cooling of the mechanical resonator within the hybrid electro-optomechanical system.
Another way to understand the cooling mechanism is through the forces exerted on the mechanical resonator. From the Hamiltonian, the force exerted on the resonator by the qubit and the cavity is given by $\partial\hat{H}/\partial x = \hbar(G\hat{\sigma}_z+G_o(\hat{a}+\hat{a}^\dagger)\,)/x_{zpf}$, where $G$ and $G_o$ are the coupling strengths and $x_{\text{zpf}}$ is the zero-point fluctuation amplitude. This force can either amplify or damp the amplitude of the resonator. For weak coupling, the rates of these processes can be computed using the Fermi's golden rule. The decay rate is given by $\Gamma_- = G^2 S_{zz}(\Omega) + G_o^2 S_{a+a^\dagger}(\Omega)$. Here, $S_{zz}(\omega)=
\int_{0}^{\infty} dt \,
e^{i\omega t}
\left\langle
\delta \hat{\sigma}_z(t)\,
\delta \hat{\sigma}_z(0)
\right\rangle, \quad \delta \hat{\sigma}_z=\hat{\sigma}_z-\langle \hat{\sigma}_z \rangle_s$
and 
$S_{a+a^\dagger}(\omega)=
\int_{0}^{\infty} dt \,
e^{i\omega t}
\left\langle
(\hat{a}+\hat{a}^\dagger)\,
(\hat{a}+\hat{a}^\dagger)
\right\rangle$ are the spectral noise densities of the qubit and the cavity at the resonator frequency $\Omega$. Conversely, the amplification rate is $\Gamma_+ = G^2 S_{zz}(-\Omega) + G_o^2 S_{a+a^\dagger}(-\Omega)$, with amplification occurring at the negative frequency of the spectral noise density. Cooling is achieved when the decay rate exceeds the amplification rate, which can be accomplished by driving the qubit and cavity at a frequency slightly less than their resonant frequencies by an amount equal to the mechanical frequency—this is known as red-detuned driving. A more detailed analysis of the cooling process can be found in \cite{PhysRevA.104.023509, marquardt_quantum_2007}.

\section{Conclusion}
\label{Conclusion}

In this review, we have presented a comprehensive overview of superconducting-circuit-based hybrid quantum systems, with a particular emphasis on their electromechanical and optomechanical realizations. 
Starting from the fundamental concepts of quantum harmonic oscillators, two-level systems, and superconductivity, we discussed the theoretical framework necessary for understanding superconducting qubits and their circuit implementations. 
We then discussed the Cooper pair box and its operating regimes, highlighting how circuit design enables the realization of charge, transmon, and fluxonium qubits, as well as their dispersive readout using microwave cavities.
Building on these foundations, we examined qubit–mechanical hybrid systems, in which mechanical resonators couple to the charge and phase degrees of freedom of superconducting qubits, particularly transmon and fluxonium qubits.
While the experimental demonstration of transmon-based mechanical coupling is well established, fluxonium-based implementations are still at an early stage, particularly due to their increased complexity. This includes the difficulty of realizing fluxonium-mechanical coupling via the flux degree of freedom. These challenges could be explored in future work.
We have discussed a wide range of phenomena and applications, including mechanical Schrodinger cat-state generation, ground-state cooling, coherent Rabi oscillations, quantum sensing, quantum state transfer, and entanglement generation.
The cat-state generation discussed here involves longitudinal coupling between the qubit and the resonator. There have also been experimental demonstrations of cat-state generation using transverse coupling via piezoelectric interactions. These cat states are typically confined within BAW resonators \cite{16microgram}. One future direction would be to implement longitudinally generated cat states in MHz nanomechanical beams.
Beyond their fundamental significance, mechanically generated Schrodinger cat states also serve as bosonic logical states for fault-tolerant quantum information processing \cite{caterror1,caterror2,caterror3,catkerr}.
The multilevel structure of anharmonic superconducting qubits, together with their coupling to mechanical resonators, may enable high-dimensional quantum information processing, with the mechanical resonator serving as a quantum bus or memory \cite{phononbus,doi:10.1126/science.aao1511,ComputationEnhancedHighDimensional}.

We further extended the discussion to qubit-mechanical-optical hybrid platforms that incorporate optomechanical interactions, thereby providing optical access to microwave quantum states. These systems offer promising pathways toward quantum transduction, long-distance quantum communication, and the realization of scalable quantum networks.
Although quantum transduction based on transverse coupling has been demonstrated and remains an active area of research, longitudinally based state-transfer protocols, where the qubit state is inferred from the coherent excitation of an optical photon, would be particularly promising.
Furthermore, there have been efforts to increase the efficiency of microwave-to-optical transduction using various schemes such as parametric amplification \cite{para_amp_trans}, feed-forward protocols for reducing added noise \cite{Higginbotham2018cn}, and entanglement-assisted transduction and teleportation of quantum signals \cite{entanglement_transduction}.
In addition to quantum transduction, hybrid electro-optomechanical systems have also been explored for quantum sensing applications, particularly for suppressing measurement-added noise in force sensing \cite{PhysRevA.111.013509}.
The thermodynamic properties of quantum systems, such as entropy production and information-theoretic aspects Landauer's principle, have been explored in optomechanical systems and trapped-ion systems \cite{Thermodynamics}. The interaction Hamiltonians employed in these studies closely resemble those discussed in this review, and such thermodynamic investigations could be extended to the hybrid systems considered here.
Overall, hybrid superconducting quantum systems constitute a versatile and rapidly advancing platform for exploring fundamental quantum phenomena and developing quantum technologies. Continued advances in device engineering, coherence enhancement, and hybrid integration are expected to further expand their capabilities and impact across quantum information science.


\subsection*{Author contributions}
Roson Nongthombam, Urmimala Dewan and  Amarendra K. Sarma are equally responsible for the conceptualization of the review, comprehensive literature research, investigation and analysis, drafting, and editing of the manuscript.

\subsection*{Funding}
This work is supported by MoE, Government of India (Grant No. MoE-STARS/STARS-2/2023-0161)

\subsection*{Data availability}
Data sharing is not applicable to this article as no new data were created or analyzed in this study.

\section*{Declarations}
\subsection*{Ethics approval and consent to participate}
Not applicable
\subsection*{Consent for publication}
Not applicable

\subsection*{Competing interests}
The authors declare no Competing interests.


\bibliography{reference1}  

\end{document}